\documentclass[journal=jctcce]{achemso}

\usepackage{amsmath}
\usepackage{amssymb}
\usepackage{bm}
\usepackage{braket}
\usepackage{caption}
\usepackage{hyperref}

\hypersetup{linktoc=all}

\setkeys{acs}{maxauthors=50}

\title{Quantum Computing for Molecular Biology}

\author{Alberto Baiardi}
\affiliation{Quantum for Life Center, Inst. f. Mol. Phys. Sci., ETH Zurich, Vladimir-Prelog-Weg 2, 8093 Zurich, Switzerland}
\email{alberto.baiardi@phys.chem.ethz.ch}
\author{Matthias Christandl}
\affiliation{Quantum for Life Center, Department of Mathematical Sciences, University of Copenhagen, Universitetsparken 5, 2100 Copenhagen, Denmark}
\email{christandl@math.ku.dk}
\author{Markus Reiher}
\email{mreiher@ethz.ch}
\affiliation{Quantum for Life Center, Inst. f. Mol. Phys. Sci., ETH Zurich, Vladimir-Prelog-Weg 2, 8093 Zurich, Switzerland}

\begin{document}

\maketitle

\begin{abstract}
Molecular biology and biochemistry interpret microscopic processes in the living world in terms of molecular structures and their interactions, which are quantum mechanical by their very nature.
Whereas the theoretical foundations of these interactions are very well established, the computational solution of the relevant quantum mechanical equations is very hard.
However, much of molecular function in biology can be understood in terms of classical mechanics, where the interactions of electrons and nuclei have been mapped onto effective classical surrogate potentials that model the interaction of atoms or even larger entities.
The simple mathematical structure of these potentials offers huge computational advantages; however, this comes at the cost that all quantum correlations and the rigorous many-particle nature of the interactions are omitted.
In this work, we discuss how quantum computation may advance the practical usefulness of the quantum foundations of molecular biology by offering computational advantages for simulations of biomolecules.
We not only discuss typical quantum mechanical problems of the electronic structure of biomolecules in this context, but also consider the dominating classical problems (such as protein folding and drug design) as well as data-driven approaches of bioinformatics and the degree to which they might become amenable to quantum simulation and quantum computation.
\end{abstract}

\section{Introduction}
\label{sec:Introduction}

The elucidation of the molecular foundations of biology has changed the way we understand biology at the cellular level.
Examples are discoveries in the molecular machinery of cells such as the structural characterization of proteins and their mode of action at atomistic resolution\cite{Drenth2007_XRayProtein-Book,Seibert2011_Virus-XRay,Chapman2011_XRay-Proteins,Spence2012_XRay-StructuralBiology} as well as the genome analysis of cells of different organisms that allowed for a correlation to the phenomenological theory of evolution.\cite{Chinwalla2002_MouseGenome,Waterson2005_ChimpanzeeGenome,Elsik2009_TaurineCattleGenome,Mardis2011_DNASequencing_Review}

Despite the fact that molecular biology addresses the atomistic resolution of biological structures and function and may therefore be expected to be inherently quantum, it is typically not perceived as a field in which quantum mechanical phenomena play a dominant role.
This is due to the fact that detailed molecular dynamics simulations rest almost exclusively on classical Newtonian mechanics. Moreover, the resulting qualitative understanding of the functioning of biomolecular machinery, supported by extensive experimental results, often does not require a detailed atomistic structure and certainly not detailed atomistic dynamics.
Instead, cartoonish representations of macromolecules are used in order to sketch a sequence of events that they undergo.
This knowledge alone is already sufficient in order to create functional networks of biochemical pathways (classical examples are the glycolysis pathway and the Krebs cycle), leading to an understanding of molecular function that may be viewed as a what-comes-next static picture of molecular events.

Still, only detailed quantitative physical simulations -- in parallel to detailed experiments with a high spatial and temporal resolution -- will allow extracting such a picture with high confidence. Classical molecular dynamics simulations provide efficient models and can be rigorously based on quantum mechanics (technically speaking, this is possible through the Born-Oppenheimer approximation,
which separates electronic and nuclear motions, with the latter
being then identified as the atomic motion in classical dynamics).
Unfortunately, more detailed simulations of the quantum mechanical equations are very difficult and only possible for a very small number of atoms.

However, if we were to advance molecular simulations by quantum computation fueled by the current hardware and algorithm developments in this field,\cite{AspuruGuzik2011_Review,AspuruGuzik2019_Review,Chan2020_Review,McArdle2020_Review,Liu2022_QuantumComputingReview} we may wonder to what extent biomolecular simulations\cite{Reiher2007_BookAtomisticBiology} would actually benefit from such developments and whether quantum computation will become key in computational quantum molecular biology.\cite{Outeiral2021,Emani2021,Fedorov2021,Marx2021}
Brought to the point, the question is, whether the emerging branches of quantum computation may eventually deliver a significant advance over traditional approaches.

\begin{figure}[htbp!]
  \centering
  \includegraphics[width=\textwidth]{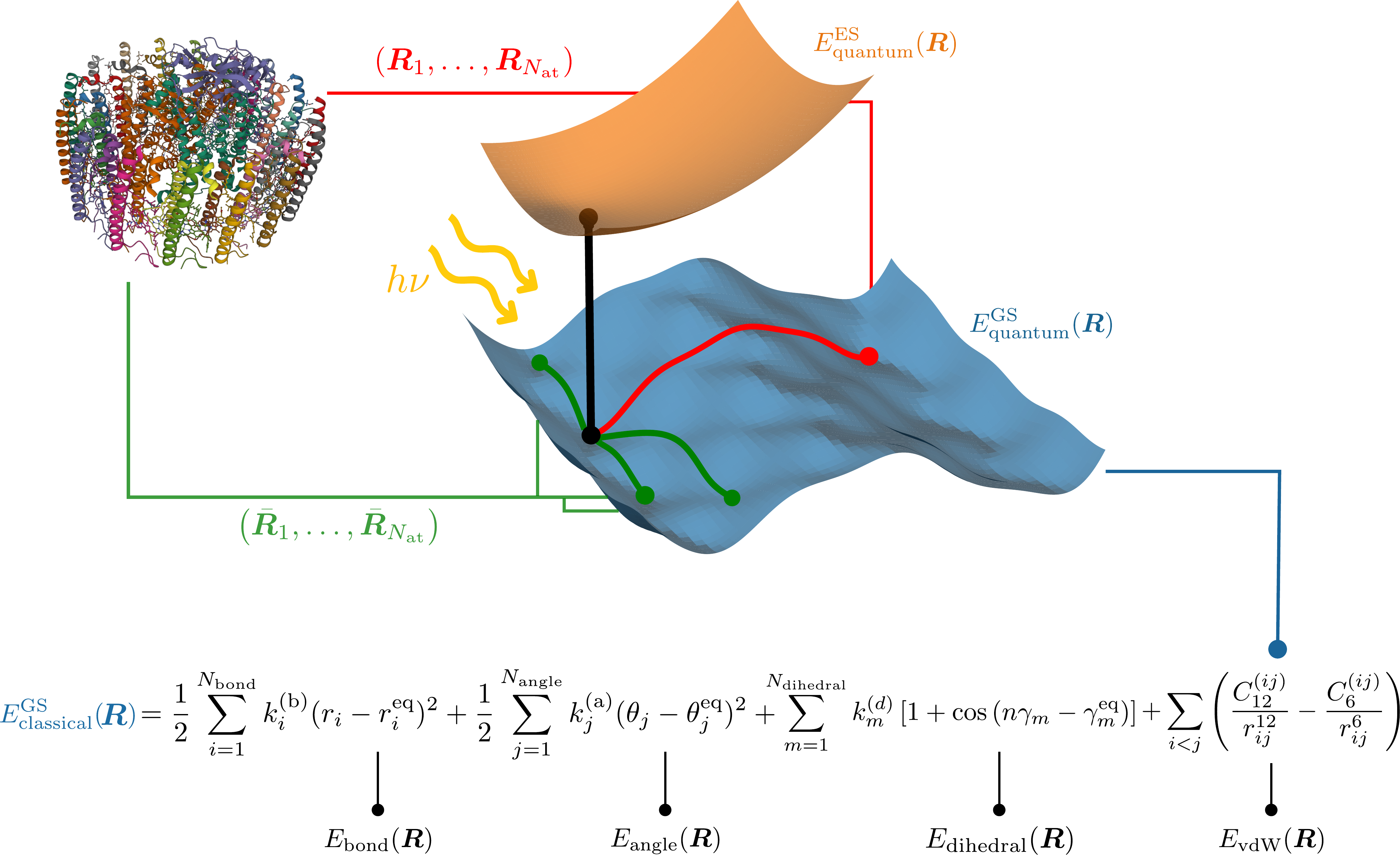}
  \caption{Graphical representation of the key steps involved in the simulation of a biological process.
  The solution of the electronic Schr\"odinger equation yields the energy of a biomolecule (e.g., the light-harvesting complex II, top left part of the picture) for a given configuration of the nuclei.
  The change of the electronic energy with the molecular structure $\left(\bm{R}_1, \bm{R}_2, \ldots \right)$ defines the potential energy surface (PES) and is represented as a blue surface for the electronic ground state.
  A biomolecule is often a large and extremely flexible entity and its PES may display many minima, which correspond to stable structures (green dots).
  Minima that are close on the PES correspond to conformers of the same molecule and are separated by a low-energy barrier that can be overcome with thermal energy.
  Minima that are well separated across the PES and by higher energy barriers (red dot) correspond to structures where the molecular connectivity changes.
  The path joining them (red line) corresponds to a reaction.
  The ground-state PES can be approximated with a classical surrogate potential $E_\text{classical}^\text{GS}(\bm{R})$ (bottom part of the figure) that is often referred to as force field.
  A force field expresses the PES, which results from the quantum interaction between electrons and nuclei, in terms of effective interactions between atoms.
  One of its parts, $E_{\text{bond}}(\bm{R})$, is associated with the bonding energy between two atoms linked by a covalent bond.
  Two further terms are added describing the variation of the energy by modifying the angle between three bonded atoms $\theta_{j}$, $E_\text{angle}(\bm{R})$, and the dihedral angle between four bonded atoms $\gamma_m$, $E_\text{dihedral}(\bm{R})$.
  Moreover, a force field will contain a term ($E_\text{vdW}(\bm{R})$) that describes the van-der-Waals interaction between pairs of non-bonded atoms that is, in turn, composed by the London attraction and the Pauli repulsion terms.
  Since the PES changes with the state of the electronic system,
  electronic excitation by resonance with an external electromagnetic field (such as light, represented as a yellow wave in the figure) switches to a new PES (we show only one of the excited electronic-state PESs in orange).}
  \label{fig:IntroductoryFigure}
\end{figure}

Whereas an intensive search is ongoing for quantum effects on function in biology,\cite{Nori2013_QuantumBiology,Plenio2014_QuantumBiologyBook,Petruccione2018_QuantumBiology,Zigmantas2020_QuantumBiology} the most important quantum effects are first and foremost rooted in the electronic structure of biomolecules and, to a lesser degree, in their quantum nuclear motion (giving rise, for instance, to tunneling and kinetic isotope effects).
The electronic structure of molecules is indeed key for the quantitative theoretical description and prediction of \emph{chemical} reactions in terms of reaction energies and activation barriers through the Born-Oppenheimer potential energy surface (PES; see Figure~\ref{fig:IntroductoryFigure}).
In other words, reaction mechanisms of chemical processes in living cells can be analyzed and understood by computationally solving the relevant quantum mechanical equations.
Such calculations provide insights into covalent bond formation in biomolecules important to biochemical pathways, into proton and electron transfer processes as well as into biological phenomena driven by visible and infrared light.

The fact that reversibility is key to molecular function in biochemistry causes macromolecules to interact predominantly through (many) non-covalent interactions.
These weak interactions, such as van der Waals forces and hydrogen bonds, can be easily broken under ambient conditions.
They originate from the quantum correlations of electrons.
However, these electron correlation effects are very hard to describe in principle in a huge many-electron system such as a protein because the computational effort for the solution of the underlying electronic Schr\"{o}dinger equation scales exponentially with system size.
This curse of dimensionality creates a steep wall that cannot be overcome with traditional computational approaches (and therefore limits such attempts to small systems of comparatively few atoms, which represent only segments of biomolecules\cite{Morokuma2006_ONIOM-QMMM-Review,Thiel2009_QMMM-Review,Ryde2016_QMMM-BookChapter}).

However, it is well known\cite{Schulten1999_Review,Schulten2001_Review,vanGunsteren2002_Review,Schulten2010_Review,vanGunsteren2013_Review} that these types of interactions can be accurately modeled by classical surrogate potentials for interacting atoms as new entities; typically only two fit parameters per pair potential are required as in Lennard-Jones potentials (note that no parameters whatsoever are required in explicit quantum chemical simulations of electronic structures).
Such force-field approximations can be combined with algorithms that, being based on classical mechanics, do not suffer from the curse of dimensionality, but have a low polynomial scaling with the number of atoms.
The dramatic gain in efficiency is a result of the mapping of the electron correlations onto pair interactions of atoms and thus mainly due to the different algebraic structure of classical vs. quantum mechanics.

Such classical model potentials are the key ingredients of biomolecular force fields in classical mechanics simulations of bio-macromolecules, which approximate the Born-Oppenheimer PES and facilitate the sampling of
the high-dimensional configuration space through classical molecular dynamics or Monte-Carlo methods in order to produce free energies from which the thermodynamic behavior of a biochemical system can be deduced.
Unfortunately, the underlying model potentials require sophisticated parametrizations, generally omit quantum mechanical effects, lack transferability to other molecule classes, and therefore cannot describe well arbitrary chemical reactions and molecular interactions (if at all).
By contrast, none of these deficiencies is present in approaches that involve quantum mechanical effects, which are, however, hampered by the highly unfavorable scaling problem with system size.

In this work, we, therefore, discuss key computational target problems in molecular biology from the perspective of quantum computation and elaborate on where and how to achieve a quantum advantage in this field.
The goal of this work is not only to deliver an overview of potential application areas and refer to pioneering results already accomplished but to give a specific roadmap on where to expect breakthroughs that would not be accessible with traditional approaches.

This paper is organized as follows: We first provide a tailored introduction to the foundations of computational physics modelling applied to molecular biology in order to establish some common ground in this  highly interdisciplinary field for a better understanding of the specific examples that follow. We begin with some principles of physical modeling of biomolecular processes (Section 2); we then consider specific quantum equations (Section 3) and their traditional solution (Section 4), before we dwell on the basics of quantum computation (Section 5).
Equipped with the key concepts and basic notation, we then turn to actual problems in molecular biology, where we first consider  the theoretical description of chemical reactions in biology (Section 6), followed by phenomena of quantum biology (Section 7) and molecular function amenable to classical molecular mechanics (Section 8).
We finish (Section 9) with remarks on the potential of data-driven methods for bioinformatics applications and eventually provide some general conclusions and outlook (Section 10).

\section{Physical modeling of biological processes}
\label{sec:PhysicalModelling}

The ways we describe, model, and understand processes in biology can be very diverse.
If we consider molecular biology as the core of biology, from which biological structures and functions, cells, and eventually the living world as a whole emerge, achievements in chemistry and molecular physics can be exploited to build a quantitative theory of the dance and interactions of biomolecules.

In this ansatz, the dynamical behavior of key entities is described by a mechanical equation of motion, which allows for a redistribution of energy among kinetic and potential contributions that eventually determine biomolecular function.
Since molecular biology rests on chemistry, it is also governed by the key theory of physics that quantitatively describes the electromagnetic interactions of electrons and atomic nuclei in and between (bio)molecules, namely quantum electrodynamics.
Starting from this basis, a quantum many-particle theory (see Section 3) and several computational solution methods (cf. Section 4) have been devised.\cite{Tannor2007_QD-Book,Sholl2009-DFTTextBook,Helgaker2013-Book,Reiher2015_RelativisticBook}
As a result, energies and molecular structures can be predicted that allow us to describe molecular behavior with breathtaking accuracy and depth.

However, what can be accomplished in terms of actual computations depends on the size of the system, i.e., on the number of electrons and atomic nuclei.
Very high accuracy can only be achieved for very small molecules.
Increasing molecular size generally is accompanied by a loss in accuracy for feasibility reasons, due to the high computational cost of exact quantum simulations.
More and more approximations need to be invoked as the molecular size grows.
\textit{Ab initio} wave-function-based methods aim at solving the many-body problem with approximate wave function parameterizations (such as truncated CI,\cite{Sherril1999_ConfigurationInteraction} coupled cluster,\cite{Bartlett2007_CC} and perturbation theory\cite{Cremer2011_MP2}), which are routinely applicable to molecules with several dozens of atoms.
Larger systems can be studied with methods that also approximate the molecular Hamiltonian in terms of parametrized models.
A prominent example is approximate density functional theory (DFT)\cite{Yang2008_DFT-Review,Sholl2009-DFTTextBook} and related approaches (for instance, the random phase approximation\cite{Ren2012_RPA}).
The model parameters are often optimized to reproduce either highly accurate quantum chemical or experimental reference data.
Although many benchmark studies have set out to quantify the accuracy of different flavors of density functional theory, it remains hard to assess \textit{a priori} the accuracy of a given density functional approximation for a specific molecule not included in the training data.\cite{Reiher2022_UncertaintyQuantification-QC}

For reasons to be discussed later, we may consider the size of a molecular system of a few hundred atomic nuclei to be routinely
treatable with approximate traditional quantum chemical approaches.
Obviously, such molecular sizes are far smaller than those of functional biological macromolecules embedded in an aqueous phase.

To alleviate this drawback, embedding techniques have been devised, which provide a quantum magnifying glass for a cluster of atomic nuclei and electrons considered to be decisive for the quantum process within a bio-macromolecule (such as the active site of an enzyme).
The neglected part of the bio-macromolecule (and of the surrounding and penetrating water molecules) is then treated as an environment.
If part of the environment will not simply play a ``spectator'' role, but is reactive -- such as a (small) part of a proteic environment (e.g., a solvent molecule or a few amino acid residues) -- it becomes part of the quantum subsystem of the embedding scheme.
The remainder of the environment is treated by a classical mechanical model with its (possibly dynamic) partial charges and geometric constraints affecting the quantum region through an additional electrostatic potential.
The determination of this quantum region can be automated (see, e.g., Refs.~\citenum{Brunken2021_AutomatedQMMM} and \citenum{Csizi2023_QMMM-Review}).
Numerous approaches have been devised for the embedding of a small active quantum part into an environment
(see, e.g., Refs.~\citenum{Tomasi2005_PCM-Review,Thiel2009_QMMM-Review,Klamt2011_COSMO-Review,Manby2012_Embedding,Mennucci2012_PCM-Review,Jacob2014_SubsystemDFT-Review,Warshel2014_NobelLecture,Wesolowski2015_Review,Mennucci2017_Embedding-LightHarvesting-Review,Muhlbach2018_QuantumSystemPartitioning,Ratner2020_EmbeddingReview,Loco2021_EmbeddingPerspective}).
We note, however, that large-amplitude motions and long-range effects may be difficult to describe in such settings.

In terms of chemical diversity, biomolecules are mostly organic molecules, often with rather regular bonding patterns (such as proteins, polysugars, nucleic acids, and other biopolymers).
It is for this reason that electronic quantum correlations turn out to be rather regular in the sense that they can be mapped to simple classical models.
These classical models are much easier to handle in actual computations, and therefore, much larger molecular sizes can be accessed (such as large proteins and protein complexes).

In other words, the underlying Coulomb interaction of electrons and nuclei within and between biomolecules can be mapped onto specific forms of potential energy functions, now defined between new dynamical entities: atoms, united atoms, or even larger coarse-grained aggregates.
Surprisingly transferable potential energy functions are obtained that capture the fundamental interactions in a molecular mechanics framework, which is now rooted in classical Newtonian mechanics rather than quantum mechanics.
Prototypical analytical forms for the distance-dependent potential energies are parametrized electrostatic and Lennard-Jones model potentials.
Only if chemical bonds are formed or broken, such descriptions become more involved and require so-called reactive force fields\cite{vanDuin2016_ReaxFF} that are, however, highly specialized and tailored towards a given class of molecules.
Moreover, their accuracy is often difficult to assess -- as is the case for any methods based on classical surrogate potentials.
Due to the lack of generality and possibility of reliably controlling the error, only explicit quantum mechanical calculations can deliver reliable insights into any biochemical process that is accompanied by breaking and forming bonds.\cite{Bertini2007_MetalloenzymesReview,Siegbahn2014_Review-Metalloenzymes,Pantazis2021_Metalloenzymes}

All of these approaches refer to the same basic concept, namely that of a potential energy hypersurface, i.e., the PES, which assigns an energy to any atomic configuration of atomic nuclei/atoms in a molecule, and hence, allows one to judge structural changes in terms of a change in potential energy. The (quantum) energies that constitute the PES are the electronic energies defined in the Born-Oppenheimer approximation which assigns energy to a fixed scaffold of atomic nuclei, in which the electrons of the molecule move.
Depending on the well depths, the number of wells, and the barrier heights exhibited by this hypersurface, different physical modeling approaches can be activated (see Figure 1).

The key take-home message is that we can describe the quantum mechanical behavior of electrons and nuclei in biomolecular systems in principle, but often not in practice, as such calculation would be too computer-time demanding.
In such cases, one resorts to classical models and accepts a reduced accuracy and a loss in transferability.
The promise of quantum computations is that a quantum description can be extended to larger atomistic structures and to higher accuracy (compared to approximate traditional quantum chemical approaches).

\section{Theoretical foundations: Quantum electronic and vibrational structure}
\label{sec:TheoreticalFoundation}

\begin{figure}[htbp!]
  \centering
  \includegraphics[width=\textwidth]{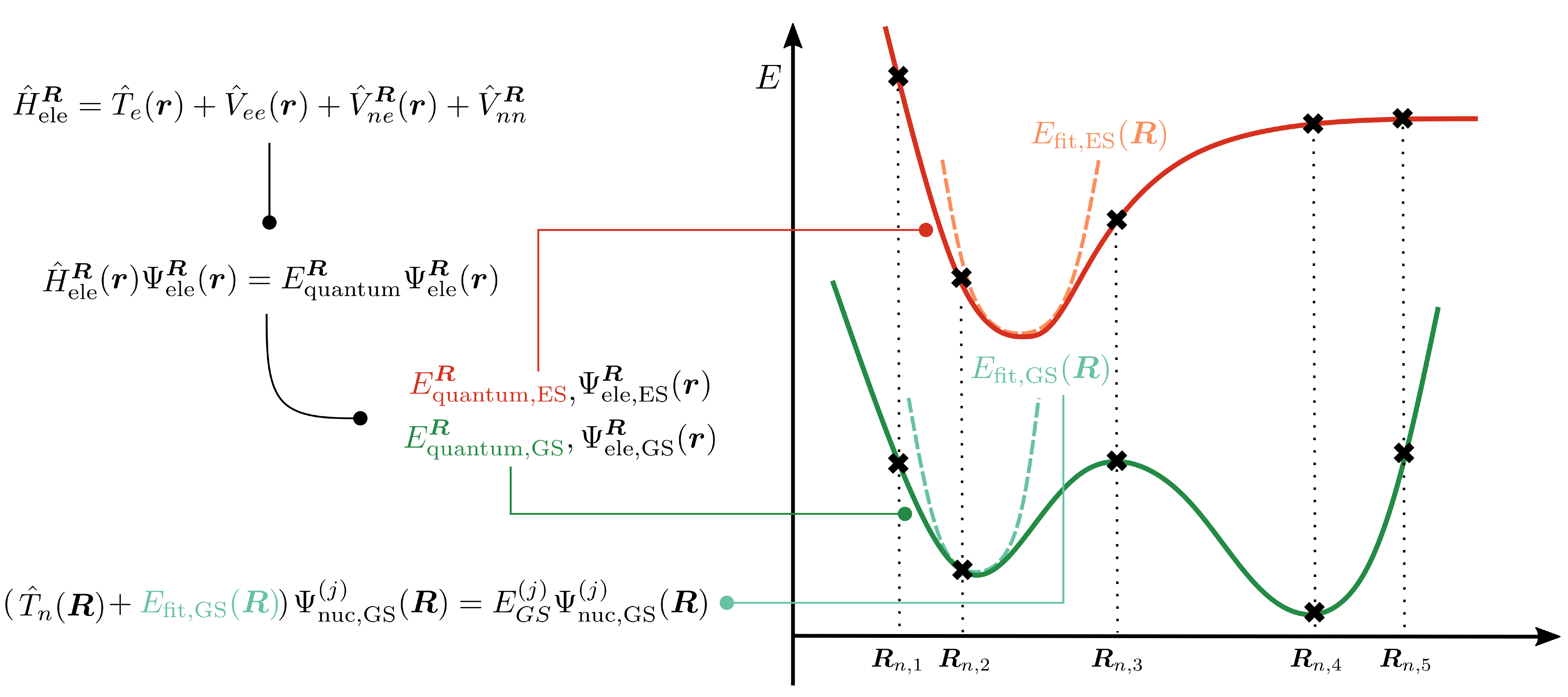}
  \caption{Graphical representation of the key equations for quantum mechanical molecular calculations.
  Electronic properties are described, for a given configuration of the nuclei $\bm{R}$, by the electronic Hamiltonian $\hat{H}_\text{ele}^{\bm{R}}$, defined in the upper left part of the figure.
  $\hat{H}_\text{ele}^{\bm{R}}$ is obtained from the molecular Hamiltonian by neglecting the nuclear kinetic energy, \textit{i.e.} by adopting the fixed-nuclei approximation.
  It contains, therefore, the kinetic energy of the electrons $\hat{T}_e(\bm{r})$, the electron-electron Coulomb interaction $\hat{V}_{ee}(\bm{r})$, the nuclear-electronic Coulomb attraction $\hat{V}_{ne}^{\bm{R}}(\bm{r})$, and the nuclear-nuclear Coulomb repulsion $\hat{V}_{nn}^{\bm{R}}$.
  $\hat{H}_\text{ele}^{\bm{R}}$ depends on the electronic coordinates $\bm{r}$ and, parametrically, on the nuclear ones $\bm{R}$, because it does not contain any derivative operator with respect to $\bm{R}$.
  The corresponding electronic ground and excited states wave functions ($\Psi_\text{ele,GS}^{\bm{R}}(\bm{r})$ and $\Psi_{ele,ES}^{\bm{R}}(\bm{r})$), together with the corresponding energies ($E_\text{quantum,GS}^{\bm{R}}(\bm{r})$ and $E_\text{quantum,ES}^{\bm{R}}(\bm{r})$) will display the same parametric dependence.
  The variation of the electronic energy of a given electronic state with the nuclear coordinates defines the PES (represented as a green curve for the electronic ground state and as a red curve for the lowest-energy excited state).
  The PES is usually only calculated point-wise (black crosses) and approximated with fitting algorithms.
  We approximate the green PES around the left local minimum by a parabolic function (the corresponding curve is displayed with a dashed line), which defines the harmonic approximation.
  The vibrational Schr\"odinger equation for the electronic ground state is reported in the lower part of the figure. Its solution yields the vibrational wave functions $\Psi_{\text{nuc,GS}}^{(j)}$.
  Since the PES changes with the electronic state, a different set of vibrational wave functions will be obtained for the excited state.}
  \label{fig:BornOppenheimer}
\end{figure}

Molecular motion is governed by the laws of quantum mechanics.
Defining these laws and the equations they lead to represent the first step toward understanding the potential of quantum computing for molecular simulations.
Quantum mechanics describes a molecule through the molecular wave function $\Psi(\bm{r}, \bm{R}, t)$, which is a complex-valued function that depends on the positions of all its elementary particles, \textit{i.e.} the electrons ($\bm{r}$) and the nuclei ($\bm{R}$), and on time $t$.
The Born rule states that the squared norm of the wave function, $\left| \Psi(\bm{r}, \bm{R}, t) \right|^2$, defines the probability density of finding the molecule in a given configuration $(\bm{r}, \bm{R})$ at a given time $t$.
The second key quantity is the molecular Hamiltonian $\hat{H}(\bm{r}, \bm{R}, t)$,
which determines the time evolution of the wave function through the time-dependent Schr\"odinger equation,

\begin{equation}
  \hat{H}(t) \Psi(\bm{r}, \bm{R}, t) = \mathrm{i} \hbar \partial_t \Psi(\bm{r}, \bm{R}, t) \, .
  \label{eq:TDSchrodingerEquation}
\end{equation}
As shown in Figure~\ref{fig:BornOppenheimer}, the Hamiltonian describing a molecule contains contributions associated with the kinetic energy of the atomic nuclei and of the electrons
of a molecular system under study,
as well as the interaction between these particles.
The interactions are very well described by pairwise electrostatic Coulomb potential energies, although they neglect, for instance, magnetic interactions.
Since these interactions do not depend on time, we
can rigorously
eliminate the $t$ dependence of the Hamiltonian, i.e., $\hat{H}(t)\rightarrow\hat{H}$.
Now, any solution to Eq.~(\ref{eq:TDSchrodingerEquation}) can be expressed in terms of specific wave functions, often referred to as \emph{stationary states}, whose probability amplitudes do not change with time.
These wave functions, $\Psi_n (\bm{r}, \bm{R})$, are the solutions to the time-independent Schr\"odinger equation,

\begin{equation}
  \hat{H} \Psi_n(\bm{r}, \bm{R}) = E_n \Psi_n(\bm{r}, \bm{R}) \, .
  \label{eq:TISchrodingerEquation}
\end{equation}
\textit{i.e.} they are the eigenvectors of the Hamiltonian $\hat{H}$ with eigenvalues $E_n$, their energy, assigned to them.

Molecular machinery in biology is viewed as intricate systems of numerous molecules, whose actions are governed mostly by reversible interactions of the biomolecules contained in the systems.
Examples are hydrated protein complexes embedded in a membrane (such as ATPase) or ribosomes.
It is because of the weakness of these reversible interactions that quantum mechanical modeling can efficiently address the quantum states of the individual molecular players (and their interactions) rather than  describing all of them as a whole system; \textit{i.e.}, separated biomolecules and their interactions are studied rather than an organelle or a cell as a whole. While this separation into components makes the solution of the Schr\"odinger equation then feasible (at least for not too large molecular structures), it also allows one to piece together and eventually understand biomolecular machinery in terms of individual energies assigned to the states of the molecular players.
For this reason, Eq.~(\ref{eq:TISchrodingerEquation}) is the main cornerstone of the simulation of biochemical processes.

Already for a molecule with a few dozen atoms, the wave function depends on hundreds of variables, and Eqs.~(\ref{eq:TDSchrodingerEquation}) and (\ref{eq:TISchrodingerEquation}) become extremely hard to solve on classical computers.
The Schr\"odinger equation can be simplified by invoking the Born--Oppenheimer approximation: since the mass of the nuclei is at least three orders of magnitude larger than that of the electrons, the nuclear dynamics are much slower than the electronic ones.
It is, therefore, reasonable to calculate the wave function in two steps.
First, the electronic wave function is calculated by neglecting the kinetic energy contribution of the nuclei.
This yields the so-called \emph{fixed-nuclei Hamiltonian} $\hat{H}_\text{ele}^{\bm{R}}$ (see Figure~\ref{fig:BornOppenheimer}), where the nuclear coordinates $\bm{R}$ do not enter as variables, but rather as fixed parameters.
Therefore, the eigenvalues $E_\text{quantum}^{\bm{R}}$ and eigenfunctions $\Psi_\text{ele}^{\bm{R}}(\bm{r})$ depend \textit{parametrically} on the nuclear coordinates $\bm{R}$, which we therefore indicate as a superscript.
They describe the allowed states and the corresponding energy of the electrons for a given configuration of the nuclei $\bm{R}$.
The nuclear part of the molecular wave function is then calculated by assuming that the change in the nuclear coordinates is slow enough to ensure that the electrons are described by the same eigenstate (e.g. the ground state) $\Psi_\text{ele}^{\bm{R}}(\bm{r})$ for all $\bm{R}$ values.
This approximation can be seen as a result of the adiabatic theorem of quantum mechanics.\cite{Born1928_AdiabaticTheorem}
The eigenvalue $E_\text{quantum}(\bm{R})$ associated with the eigenfunction describing the electrons defines the effective interaction potential between the electrons and the nuclei.
The nuclear wave function $ \Psi_\text{nuc}(\bm{R})$ describing this motion is, therefore, obtained by solving the Schr\"odinger equation for a Hamiltonian containing $E_\text{quantum}(\bm{R})$ as interaction term (see Figure.~(\ref{fig:BornOppenheimer})).
The overall molecular wave function is then given by $\Psi_\text{nuc}(\bm{R}) \times \Psi_\text{ele}^{\bm{R}}(\bm{r})$.

Whether one needs to solve either the time-independent or the time-dependent Schr\"odinger equation, either only for the electrons or also for the nuclei, depends on the biochemical process at hand.
In all cases, the wave function is the solution of a differential equation in which all variables are coupled by the potential energy operator.
This coupling correlates the motion of the quantum particles and renders the solution of the differential equation a daunting computational task for large molecules.

\begin{figure}[htbp!]
  \centering
  \includegraphics[width=.8\textwidth]{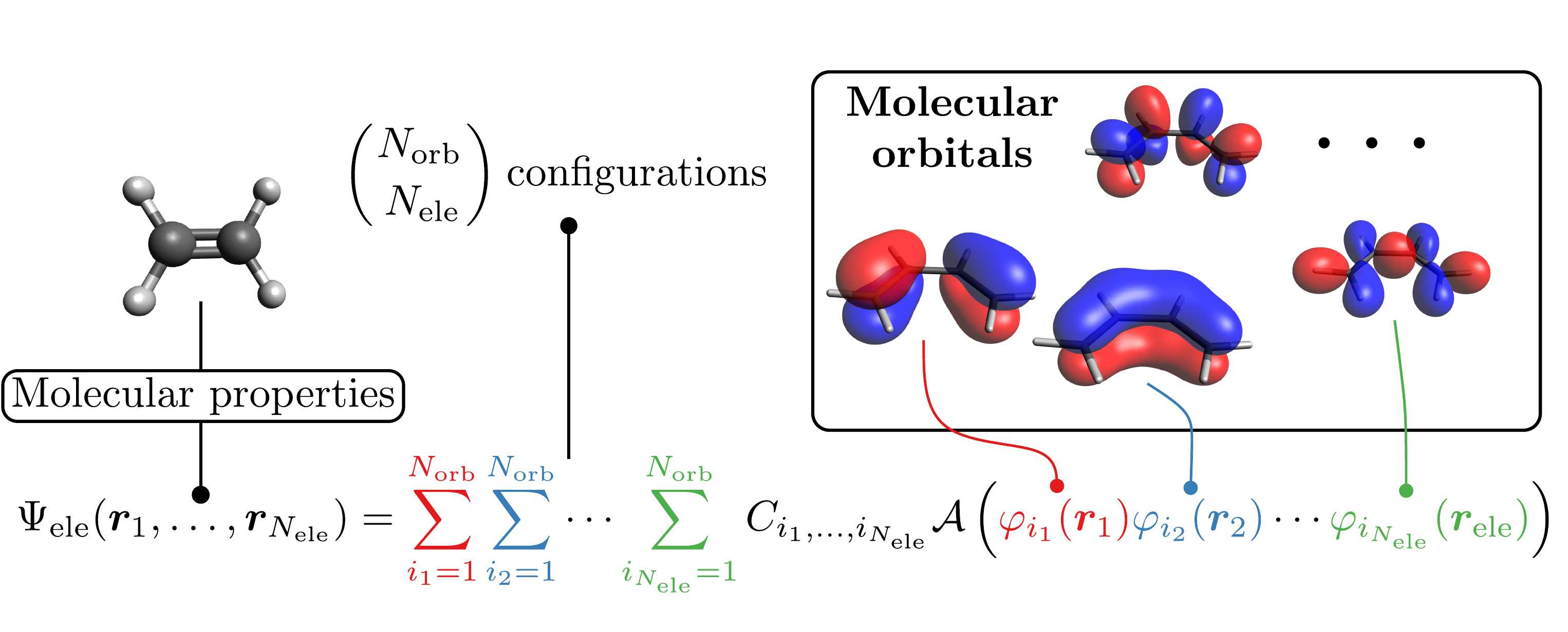}
  \caption{Graphical representation of how the full-CI wave function is constructed starting from a given set of molecular orbitals for the butadiene molecule.
  The full-CI wave function $\Psi_\text{ele}(\bm{r}_1, \ldots, \bm{r}_{N_\text{ele}})$ is expressed as a linear combination of antisymmetrized products of the molecular orbitals $\varphi_i(\bm{r})$ (a graphical representation of the set of molecular orbitals of ethylene is given in the right upper part in blue and red, indicating the different signs of these functions).
  From each molecular orbital, two spin orbitals, one for spin-up and one for spin-down electrons, are constructed.
  The antisymmetrization operator $\mathcal{A}$ is applied to each configuration to ensure that each of these basis functions is antisymmetric upon exchange of electrons -- as prescribed by the Fermionic nature of the electrons.
  The coefficients entering the linear combination are stored in the tensor $C_{i_1,\ldots,i_{N_\text{ele}}}$, which has one index per electron.
  By virtue of the Pauli exclusion principle, a spin orbital can be occupied by at most one electron.
  Therefore, as shown in the upper part of the figure, the overall number of terms included in the linear combination is equal to the number of subsets of $N_\text{ele}$ elements of the molecular orbital set, which contains $N_\text{orb}$ elements.
  This binomial scaling results in the curse of dimensionality.}
  \label{fig:CIWaveFunction}
\end{figure}

Why is the solution of many-body quantum problems such a major computational hurdle?
We consider the electronic problem, although our conclusions will apply to the nuclear one as well.
In order to represent it on hardware, either classical or quantum, the many-electron wave function must be discretized.
This discretization is realized by expressing the wave function in terms of a finite set of one-electron functions $\left\{ \varphi_k(\bm{r}, \sigma) \right\}_{k=1}^{N_\text{orb}}$, the so-called \emph{molecular orbitals}.
Specifically, the wave function is expressed as a linear combination of antisymmetrized products of molecular orbitals as sketched in Figure~\ref{fig:CIWaveFunction}.
An antisymmetrized product is often referred to as \emph{Slater determinant} or as \emph{electronic configuration}.
Methods that solve the Schr\"odinger equation within the resulting wave function ansatz without any restriction of the space of configurations are, in chemistry, referred to as \emph{full configuration interaction} (full-CI), because the wave function is expressed as a linear superposition of all electronic configurations. In physics, this is known as \emph{exact diagonalization}.
The molecular orbitals are themselves discretized as a linear combination of a finite set of orbitals -- usually referred to as \emph{atomic orbitals} to distinguish them from the molecular ones.
The atomic orbitals are basis functions that are pre-optimized to represent the orbitals of a given atom (and its polarization) accurately.
They are expressed, in practice, as a linear combination of Gaussian basis functions.
This facilitates the calculation of integrals that define the representation of the molecular Hamiltonian in the atomic orbital basis.

For a molecular system, the union of the atomic orbitals of all atoms defines the atomic basis in which the molecular orbitals are expanded.
The expansion coefficients are called molecular orbital coefficients, which can then be optimized based on the variational principle.
When the optimization is only carried out for a single Slater determinant, this yields the Hartree-Fock (HF) method.

In the limit of a dense, complete atomic basis set -- a limit called the \emph{complete basis set limit} -- the many-electron wave function can be encoded exactly as a linear superposition of Slater determinants.
The complete basis set limit cannot be reached in numerical simulations that must necessarily rely on finite, incomplete basis sets.
Extrapolation techniques have been designed to tame this problem.
These methods assume that the electronic energy $E_\text{ele}$ depends on the basis set size $n$ according to a given, fixed power law.
One can then solve the electronic Schr\"odinger equation for increasingly larger basis sets, determine the power law coefficients by least-squares fitting, and estimate the complete basis set electronic energy $E_\text{ele}$ from the limiting value obtained for $n \rightarrow +\infty$.
These methods are, however, inherently heuristic because the form of the power law must be fixed \textit{a priori}.

Encoding an extremely complex high-dimensional wave function in terms of much simpler entities, the orbitals, comes with a price.
Let us consider a molecule with $N_\text{ele}$ electrons described by $N_\text{orb}$ molecular orbitals.
Let us also assume that $N_\text{orb}$ is proportional to $N_\text{ele}$.
This is a reasonable assumption because, as highlighted above, $N_\text{orb}$ is proportional to the number of atoms of the molecule that is, in turn, proportional to $N_\text{ele}$.
Since one can choose any one of the $N_\text{orb}$ molecular orbitals to describe each electron, the number of configurations that can be constructed grows combinatorially with $N_\text{ele}$.
This implies that the computational representation of the wave function requires storing a number of coefficients that grows exponentially with the number of electrons.
The resulting unfavorable scaling is often referred to as the \emph{curse of dimensionality} and renders full-CI impractical on classical hardware already for molecular systems with more than three to five atoms.

The curse of dimensionality plagues any full-CI-based method aiming at solving quantum many-body problems in terms of one-particle functions.
This hurdle is, therefore, not limited to time-independent electronic calculations.
The same challenges will be faced for the solution of the time-dependent Schr\"odinger equation, both for the nuclear and for the electronic case.
The form of the differential equation changes -- the time-independent Schr\"odinger equation is an eigenvalue equation, while the time-dependent one is a partial differential equation that involves both the spatial and the time variables.
Still, both equations are defined in terms of the Hamiltonian operator that couples the variables on which the wave function depends.
Full-CI is, therefore, needed to describe the resulting quantum correlations completely, leading once more to the emergence of the curse of dimensionality.

Research in quantum chemistry has mostly focused on taming the curse of dimensionality with algorithms that trade the full-CI accuracy and generality for computational efficiency.
Specifically, they rely on wave function approximations which scale polynomially in the number of expansion parameters but are tailored to a specific regime of \emph{electron correlation}.
In quantum chemistry, the term \emph{electron correlation} denotes an ensemble of effects that are not accounted for by the Hartree--Fock ansatz. This notion corresponds well with notions of correlation and entanglement in the context of quantum information theory, the number of necessary configurations corresponding to the Slater rank\cite{Cirac2001_SlaterRank} -- the Fermionic analog of the Schmidt rank quantifying quantum entanglement.\cite{NielsenChuangBook}
More detailed information about the weight components can be cast as eigenvalues of the one-particle reduced density matrix with naturally associated entropic measures and relation to Pauli's principle.\cite{Klyachko2008_GeneralizedPauli,Reiher2012_Entanglement-Correlation,Stein2016_AutoCAS,Schilling2018_GeneralizedPauli}

Consider the idealized case in which the Coulomb repulsion between the electrons is vanishing and, therefore, the electrons are not coupled.
In this case, the full-CI wave function coincides with the Hartree-Fock determinant.
Therefore, electron correlation effects are those that impede encoding the wave function as a single determinant and require it to be expressed as a linear superposition of
configurations.

Qualitatively speaking, the larger the number of configurations with non-negligible expansion coefficients, the stronger the correlation effects will be.
Paradigmatic examples of strong correlation effects in molecules are transition metal complexes (as they occur in active sites of metalloenzymes and FeS clusters) and bond-breaking processes.\cite{Baiardi2020_Review,Shiozaki2020_Review,Roemelt2021_Review}

In the former case, the manifold of $d$-type atomic orbitals of transition-metal active sites, and transition metal complexes in general, often contains near-degenerate orbitals that are partially occupied.
The corresponding configurations will, therefore, have a similar weight in the full-CI wave function.
Similarly, bond-breaking processes yield electronic configurations where unpaired electrons are localized on different parts of the molecule -- a situation that cannot be captured by a single configuration in a singlet state because spin pairing will always produce two degenerate configurations with alpha and beta spin orbitals exchanged on the dissociating parts.

Moreover, weak electron correlation effects -- referred to as dynamical correlations in quantum chemistry -- must also be taken into account.
Although they do not alter the qualitative structure of the full-CI wave function, they yield a non-negligible contribution to the electronic energy because of the sheer number of configurations with very small weights.

Dynamical correlations comprise two classes, short- and long-range correlations. From a physical perspective,
long-range correlations are associated with the London
dispersion forces, which, qualitatively speaking, describe the induced dipolar interaction between two molecular fragments that result from instantaneous polarization of the corresponding electronic density.
Although they require, in principle, a proper quantum description of the electron-electron interaction, dispersion interactions can be well-approximated with effective pairwise classical potentials.\cite{Grimme2011_Review}

Short-range correlations are, instead, associated with the short-range electron-electron interaction, \textit{i.e.}, in the limit of a vanishing interelectronic distance in which the positions of two electrons coalesce.
It can be proven that, at the coalescence of two electrons,
the many-body wave function displays a cusp (the so-called Kato cusp), \textit{i.e.} its derivative is not continuous.
Attempting to fulfill the Kato cusp condition with products of cuspless Gaussian basis functions yields highly non-compact full-CI wave functions, with many configurations with small but non-negligible coefficients -- the signature of dynamical correlations. Accordingly, the basis set expansion of the wave function is often supplemented by factors that depend explicitly on the interelectronic distance\cite{Valeev2012_F12-Review}

\section{Computational foundations: Traditional approaches}
\label{sec:TraditionalApproaches}

\noindent The curse of dimensionality limits practical applications of full-CI to molecules with approximately 20 electrons,\cite{Vogiatzis2017_MassivelyParallelDMRG,Baiardi2020_Review} a size that characterizes very small molecules, far smaller than any biomolecule.
The impossibility of routinely applying full-CI to large molecules has sparked the development of approximate methods not suffering from the curse of dimensionality.
The key idea behind many of these approximations is to identify wave function parametrizations that can approximate full-CI while being defined by a number of parameters that scales polynomially with the system size.
These approximate parametrizations have been tailored to two idealized situations: weakly correlated and strongly correlated electronic structures of molecules.
Unfortunately, strong and weak correlation effects often coexist in a molecular system and, therefore, these methods must be combined in accurate molecular simulations.

\begin{figure}[htbp!]
  \centering
  \includegraphics[width=\textwidth]{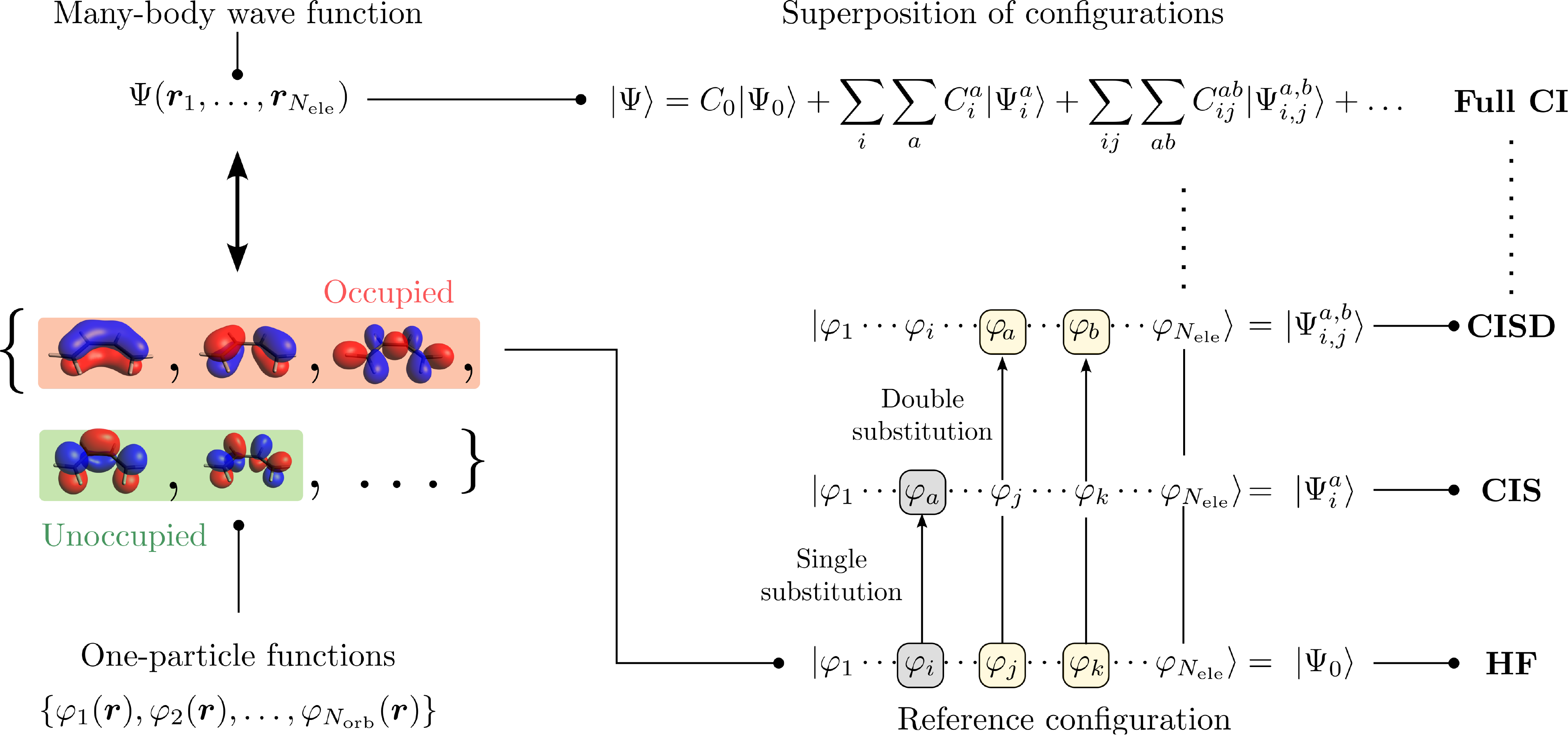}
  \caption{Graphical representation of configuration interaction-based wave function approximations (illustrated for the electronic wave function of butadiene).
  Configuration Interaction methods encode the many-body wave function (upper part of the figure) in terms of one-particle functions, known as orbitals in electronic structure theory (bottom left part of the figure).
  Molecular orbitals may be optimized with Hartree-Fock or Density Functional Theory and are expressed as a linear combination of a $N_\text{orb}$-dimensional basis set.
  Out of the $N_\text{orb}$ molecular orbitals obtained from such a calculation, the $N_\text{ele}/2$ ones that yield the lowest-energy determinant are called occupied and the other ones are denoted virtual.
  The occupied orbitals define the reference determinant (a determinant is often referred to as configuration).
  The configurations entering the full-CI wave function can be classified in terms of the number of orbitals by which they differ compared to the reference determinant.
  Configurations differing by a single spinorbital are referred to as 'single excitations' (represented in gray), while the ones differing by two spin orbitals are referred to as 'double excitations' (represented in yellow).
  Truncated CI approximates the full-CI wave function by including only all those configurations up to $N_\text{sub}$-fold excitations.
  For $N_\text{sub}=1$, one obtains the configuration interaction singles (CIS) approximation, whereas for $N_\text{sub}=2$ the method is referred to as configuration interaction with singles and doubles excitations (CISD).
  If $N_\text{sub}$ coincides with the overall number of electrons of the molecule, one retrieves full-CI.
  Note that the general notation for the expansion coefficients $C_{i_1,\ldots,i_{N_\text{ele}}}$ in Figure~\ref{fig:CIWaveFunction} is here replaced to follow the partitioning according to excitation classes as $C_i^a$, $C_{ij}^{ab}$, ...}
  \label{fig:FullCI}
\end{figure}

Above, we defined weakly correlated wave functions as those in which the Hartree--Fock configuration has a predominant weight (given as the square of its expansion coefficient in the expansion of configurations).
Since the wave function is normalized to a fixed value (typical one), all other determinants will have a comparatively small weight.
Moreover, it is reasonable to expect that the weight of the other configurations will decrease as their departure from the Hartree-Fock determinant increases.
This departure can be measured in terms of so-called \emph{excitations}, which are orbital substitutions (see Figure~\ref{fig:FullCI}): a determinant will be defined as a $N_\text{sub}$-fold excited determinant if it differs from the Hartree-Fock reference one by at most $N_\text{sub}$ orbitals.
We note that the term \emph{excitation} is misleading, despite having become a standard notion, as it implies that an orbital substitution corresponds to an electronically excited state, which is only true for non-interacting electrons and obvious when considering that the `excitations' are still elements of a many-electron determinantal basis set into which the ground-state wave function is expanded in the configuration interaction ansatz.
Based on this idea of orbital substitutions, the full-CI wave function can be systematically approximated by including only up to $N_\text{sub}$-tuple substitutions in the wave function expansion.
This leads to so-called truncated CI approaches, whose computational complexity scales polynomially with the number of orbitals, with an exponent proportional to $N_\text{sub}$.
We note that the emerging hierarchy of approximations is an ad hoc choice that does not guarantee that all truly important configurations are actually included in the truncated expansion of configurations.
The key limitation of truncated CI (by contrast to full-CI), however, is that it is not size extensive, \textit{i.e.} the corresponding energy does not scale linearly with the number of electrons,\cite{Shavitt1998_CI-Review} which makes it impossible to reliably compare the energy of molecules with different numbers of electrons.

The established canonical solution to this problem is the coupled cluster (CC) method\cite{Musial2007_Review}.
CC relies on an exponential wave function ansatz that contains a number of parameters that scales polynomially with system size.
Specifically, CC with singles, double, and perturbative triple excitations, abbreviated CCSD(T), has become the \textit{de facto} reference method for calculating weakly correlated electronic wave functions in order to reach chemical accuracy of about 1 kcal/mol.

Approaches such as truncated CI or CC -- usually referred to as \emph{single-reference} methods, because they rely on a single configuration, the Hartree--Fock one, as a reference -- can become inaccurate as soon as the full-CI wave function is not dominated by a single configuration.
For such multi-reference cases,
modern full-CI-type methods\cite{Baiardi2020_Review,Eriksen2021_FCI-Review} have made it possible to push the limits of full-CI to systems with approximately 100 electrons.
These methods can be qualitatively divided into two classes: selected configuration interaction methods and algorithms based on tensor network states.

Selected configuration interaction methods rely on the assumption that, although not being dominated by a single configuration, the full-CI expansion remains highly sparse so that the number of negligible expansion coefficients $C_{i_1,\ldots,i_{N_\text{ele}}}$ (see Figure~\ref{fig:CIWaveFunction}) is in overwhelming majority.
Full-CI Quantum Monte Carlo (FCIQMC)\cite{Alavi2009_FCIQMC} exploits this idea by solving the Schr\"odinger equation \textit{via} a stochastic Monte Carlo-based realization of the imaginary-time evolution algorithm.
Deterministic variants of FCIQMC are also available,\cite{Neese2003_SORCI,Holmes2016_HBCI,Zimmerman2017_IncrementalCI,Schriber2017_SelectedCI,Mayhall2020_SelectedCI-TensorProduct} which originate from the ``Configuration Interaction using a Perturbative Selection made Iteratively'' (CIPSI) method introduced by Malrieu\cite{Malrieu1973_CIPSI} that identifies the non-negligible configurations based on second-order perturbation theory.
In practice, selected CI methods can target up to 100 electrons, which is a huge leap forward for chemical applications, but still represents a severe limitation for generally large biomolecules.
Surely, other approximate quantum approaches that do not rely on expansions of configurations -- most importantly contemporary methods of Density Functional Theory (DFT; see below) -- can target systems that are two orders of magnitude larger, but they also come with an uncontrollable error for a specific biomolecule under consideration.\cite{Reiher2022_UncertaintyQuantification-QC}

An alternative route is offered by tensor network states (TNSs).\cite{szalay2015tensor}
In essence, a tensor network approximates the tensor $\bm{C} = C_{i_1,\ldots,i_{N_\text{ele}}}$ of all full-CI expansion coefficients (cf. Figure~\ref{fig:CIWaveFunction}) with tensor compression algorithms.
This tensor compression reduces the number of wave function parameters but is inevitably associated with a loss of accuracy.
However, TNSs can be designed to yield an optimal balance between compression degree and accuracy.
In many numerical studies it turned out that a specific class of TNSs, so-called matrix product states, reliably approximate the electronic Schr\"odinger equation,\cite{Chan2008_Review,Zgid2009_Review,Marti2010_Review-DMRG,Schollwoeck2011_Review,chan2011density,Wouters2013_Review,Kurashige2014_Review,Olivares2015_DMRGInPractice,szalay2015tensor,Yanai2015,Baiardi2020_Review} whose parameters can be very efficiently optimized with an iterative algorithm known as density matrix renormalization group (DMRG) approach.\cite{white92,white93}
Although we have focused so far on the key problem of molecular quantum simulations, namely the solution of the time-independent electronic Schr\"odinger equation (cf. Figure~\ref{fig:BornOppenheimer}),
we note that the approaches discussed so far can be generalized to the nuclear problem as well (see, for example, Ref.~\citenum{Baiardi2017_vDMRG} for the introduction of vibrational DMRG) and to the time domain to solve time-dependent Schr\"odinger equations.\cite{Paeckel2019_TDDMRG,Baiardi2019_TDDMRG}

Even for complex molecules with several dozen atoms, electronic correlation is rarely strong for more than 100 electrons.
This has inspired the development of so-called active-space algorithms: a full-CI-based scheme is applied only on the strongly correlated orbitals, while the remaining weak correlation effects are captured \textit{a posteriori}.
Translating such an apparently simple idea into an algorithm is a hard task.
In fact, designing algorithms to include accurately weak correlation effects on top of a DMRG or FCIQMC simulation remains an open challenge of computational chemistry.
One should not forget that the second hurdle of active space-based methods is the need of identifying the strongly-correlated electrons \textit{a priori}, without knowing the exact full-CI wave function.
In this respect, various strategies have been proposed to efficiently identify strong correlation effects based either on single-reference calculations or on partially converged full CI-type simulations.\cite{Pulay1988_NOON-ActiveSpace,Veryazov2011_ActiveSpaceSelection,Stein2016_AutoCAS,Chan2017_AVAS,Stein2019_AutoCAS-Program}

Finally, we emphasize that DFT represents a route to solve the electronic structure problem that has become \emph{the} workhorse of computational molecular quantum mechanics.
The key difference between a superposition of configurations and DFT is that
DFT calculations in practice rely on the Kohn-Sham variant, where the density $\rho(\bm{r})$ is obtained from a single determinant wave function in such a way that it can be calculated as a sum of absolute squares of the orbitals entering the determinant.
The key insight here is that one can obtain the exact ground-state electron density of an interacting quantum
system from this determinant, which describes exactly a non-interacting surrogate system of fermions.
The orbitals are obtained by solving a Hartree-Fock-like set of equations that is defined in terms of the so-called exact
density functional $E[\rho(\bm{r})]$.
While this procedure is exact, the exact form of $E[\rho]$ is, unfortunately, not known.
Practical DFT calculations rely on heuristic forms of $E[\rho]$ that are either data-driven or inspired by the form of the functional obtained for simple model systems.
Estimating the accuracy of approximate functionals is often impossible,\cite{Reiher2022_UncertaintyQuantification-QC}  which prevents quantifying and systematically improving the accuracy of a DFT calculation, which is the major drawback of contemporary DFT.

We pause at this point for the conceptual observation that the ground state energy can be written in variational terms as
\begin{equation}
  E_g = \min \bra{\psi} H \ket{\psi}
  \label{eq:MinimzationProblem}
\end{equation}
and that methods that optimize over an ansatz (variational methods) such as Hartree-Fock, full-CI, DFT, restrict the minimization and always lead only to an upper bound on the true ground state energy.
Whereas efforts can be made to assess the closeness to the true ground state energy (e.g. extrapolation to an infinite basis set in the case of full-CI), there can ultimately not be any rigorous closeness estimate of such methods (as the infinite basis set limit cannot be reached in practice) or the ansatz within a given basis set only samples a subset of the states (e.g. Hartree-Fock, CC, ...).

It was already noted in 1960 by Coulson\cite{coulson-1960}
that rigorous lower bounds could be obtained by the observation that molecular Hamiltonians only involve pairwise interactions $H=\sum_{ij} h^{(2)}_{ij}$ and that, therefore, the ground state energy can be cast as an optimization over the two-particle reduced density matrix (2-RDM) $\rho^{(2)}$

\begin{equation}
  E_g= \binom{N}{2} \min \text{Tr} \left( h^{(2)} \rho^{(2)} \right).
  \label{eq:GroundStateEnergy}
\end{equation}

The function to be optimized is linear and the set of 2-RDMs is convex and only of polynomial dimension in the basis set, sparking hope for a solution to the problem. The earlier discussed difficulty of the ground state energy problem is now hidden in the complexity of describing this convex set. The concrete question is: Given a matrix $\rho^{(2)}$, is it the reduced density matrix of an $N$-particle electronic wave function?
This problem is known as the N-representability problem of quantum chemistry
and solving it as Coulson's challenge.\cite{tredhold-1957, coulson-1960, coleman-RMP}.

Whereas the exact characterization of the set has been shown to be Quantum Merlin Arthur (QMA)-hard (QMA is the quantum analog of the famous class of computational problems known as nondeterministic polynomial time (NP)) and thus unlikely to be solved even on a quantum computer \cite{QMA-complete}, we note that any \emph{outer} approximation to the set will result in a \emph{lower} bound on the energy, in stark contrast to the situation discussed above for the variational methods.

Concretely, such approximations can be given by semi-definite constraints (rendering the optimization problem a semi-definite program) on the set and a number of nontrivial constraints have been found and, in general, significant efforts have been devoted to this problem. \cite{coleman-yukalov, Cioslowski, mazziotti-2006} For chemical problems, unfortunately, the rigorous lower bounds obtained with this method have not been as close from below to the believed true ground state energy as the variational methods from above.
The success of such methods has therefore been limited in practice.

Many of the methods outlined so far target the lowest-energy solution of the electronic Schr\"odinger equation
(see Figure.~(\ref{fig:BornOppenheimer}))
which corresponds to the electronic ground state.
The ground-state energy is the key quantity for reactions that happen on the electronic ground-state PES (see Figure~\ref{fig:IntroductoryFigure}).
However, some biological processes are driven by the absorption of electromagnetic radiation.
In this case, the reaction involves the surface of an electronically excited state, as shown in Figure~\ref{fig:IntroductoryFigure}, which corresponds to higher-energy solutions of Eq.~(\ref{eq:TISchrodingerEquation}). These solutions are not only important to describe a biomolecular system after light excitation, but they also offer information that can be related to spectroscopic techniques and, hence, to experimental spectroscopies that identify specific states of biomolecular machinery.
Many of the methods reviewed above can be straightforwardly generalized to target excited states -- although these generalizations are often harder than their ground-state counterparts.

Molecules can be subjected to spectroscopic techniques through their response to external electromagnetic perturbations (i.e., light).
The spectroscopic response of a molecule does not depend only on its energy,
but also on its properties, such as the electric and magnetic dipole moment.
Response properties can be calculated based on the methods listed above with time-dependent perturbation theory.\cite{Helgaker2012_MolecularProperties}

Identifying the method providing the best compromise between cost and accuracy for time-dependent simulations is an especially hard task.
When driven out of its ground state, the nature of the wave function and, therefore, the correlation between its particles may change, even dramatically.
It is known that the time evolution of a wave function following an external perturbation (usually known as \emph{quench} in quantum physics, which corresponds most often to an electromagnetic perturbation in chemistry) may lead to a continuous growth of its entanglement.
This makes it difficult to accurately represent the time-evolving wave function as a TNS over the whole propagation.
For example, the entanglement growth makes the MPS representation of the wave function increasingly inaccurate with increasing time.
How this loss in accuracy affects the simulation accuracy depends on the simulation target at hand.

\section{Computational foundations: The promise of quantum simulation}
\label{sec:quantum}

It was Feynman\cite{Feynman1982_QuantumSimulation} who noticed that using a classical system, e.g. a traditional computer, to simulate a quantum system will in general result in an exponential overhead in the time required for the simulation due to the curse of dimensionality, but that building a simulator (or computer) directly from quantum mechanical constituents (henceforth called a quantum simulator or, if very versatile, quantum computer) could circumvent this curse and therefore achieve an efficient simulation.
With efficient simulation we mean a simulation that requires time and space resources that scale proportionally (or at most polynomially) with the size of the biomolecule
to be simulated.
Analogous to the bits of a computer, the constituents in a quantum computer are most naturally cast in terms of quantum bits (qubits), quantum mechanical two-level systems.
Qubits in a quantum computer are combined with the tensor product operation.
The idea to build a quantum system to simulate another quantum system --- that \emph{quantum simulates quantum} --- is as self-evident and simple as it is revolutionary.

The technical reason that hinders classical simulators at the possibility of simulating quantum systems is the different composition rules of its elementary components.
In fact, the classical composition rule is weaker than the quantum mechanical.
This means, in practice, that only a subset of the possible quantum mechanical states of $N$ qubits can be represented by $N$ of their classical counterparts.
This is precisely remedied by building a simulator directly from quantum mechanical constituents, since those evidently adhere to the same quantum mechanical composition rule and, therefore, offer a computational platform perfectly suited to the task of simulating a quantum system.

By simulating a quantum system we mean, just as before, simulating relevant properties of quantum systems -- the most relevant being, in chemistry and biology -- molecular properties.
Primary examples of relevant quantities to be simulated are, for static phenomena,  ground- and excited-state energies and, for non-equilibrium processes, time-dependent correlation functions.
Simulating quantum systems is not the unique application of quantum computing.
Other computational problems that are relevant for physics, chemistry or biology and that are hard to solve on a classical computer may be solved more efficiently on quantum computers.
A candidate in the biological context is the protein folding problem, which is an optimization problem,  for which heuristic quantum optimization algorithms are being investigated.\cite{Fingerhuth2018,Robert2021}

It is important to emphasize that, although \emph{quantum simulates quantum} is such a self-evident slogan, the existing quantum simulation algorithms are highly non-trivial and problem-dependent.
Their future development will therefore require ingenuity and insight both from the general principles of quantum algorithmic design (mathematics, physics, and computer science) and from the application side (physics, chemistry, biology).

Quantum simulators naturally fall into two categories, traditionally called \emph{analog} and \emph{digital} quantum simulators.
\emph{Analog quantum simulators} are more akin to purpose-built machines that through their build-up mimic the interactions and, therefore, the behavior
of the system to be simulated.
When equipped with knobs to modify the behavior of the simulator they can have aspects of universality.\cite{lloyd-1996,cubitt-pnas-2018}
Yet, they are meant as purpose-specific computing machines and, therefore, are not meant to be universal computing machines.
Mimicking the interactions of a quantum system with an analog quantum simulation is often hard due to the fact that the many-body interaction terms of a given Hamiltonian (for instance, the electronic Hamiltonian of Figure~\ref{fig:BornOppenheimer})
may not be directly encoded on the hardware at hand.
This limit can be overcome with so-called gadgets that map, via perturbation theory, the desired interactions of the system to be simulated to effective interactions built from those available in the hardware.
The price to be paid is a slight increase, usually a multiplicative constant, of the hardware size needed to simulate a given quantum system.

The amazing experimental progress, especially in the area of optical lattices, where many atoms are trapped and interact in a lattice of standing light waves, has delivered remarkable simulation results of condensed matter systems surpassing the practical simulation capability of traditional computers\cite{cirac-2012,dayley-2022}.
Yet, since the constituents of analog quantum simulators also experience decoherence --- a quantum form of errors --- which cannot be suppressed arbitrarily (this is true at present, and may remain so even in the far future), formal guaranteed statements about the overall accuracy of the achieved simulation are difficult if not impossible.
Control in such a system is continuously being improved, yet, in general, experimental analog quantum simulators are characterized by limited control and accuracy, but have a relatively easily scalable number of constituents.
The limited control constitutes a major challenge when simulating (bio)chemical processes with digital simulators.
The key difference between Hamiltonians relevant in quantum physical problems and those appearing in quantum chemical and quantum biological contexts is that the (discretized)
interaction is drastically more irregular in the latter case than in the former one.
Tunability is, therefore, a key requirement of an analog quantum simulator for quantum chemical and quantum biological simulations.

In a nutshell, analog quantum simulators are single-purpose machines with an easily scalable size designed to mimic the behavior of interesting quantum systems.
Affected by decoherence on the individual-component level and without complete error-correcting and fault-tolerant mechanisms in place, however, few rigorous computational guarantees are available.

\begin{figure}[htbp!]
  \centering
  \includegraphics[width=\textwidth]{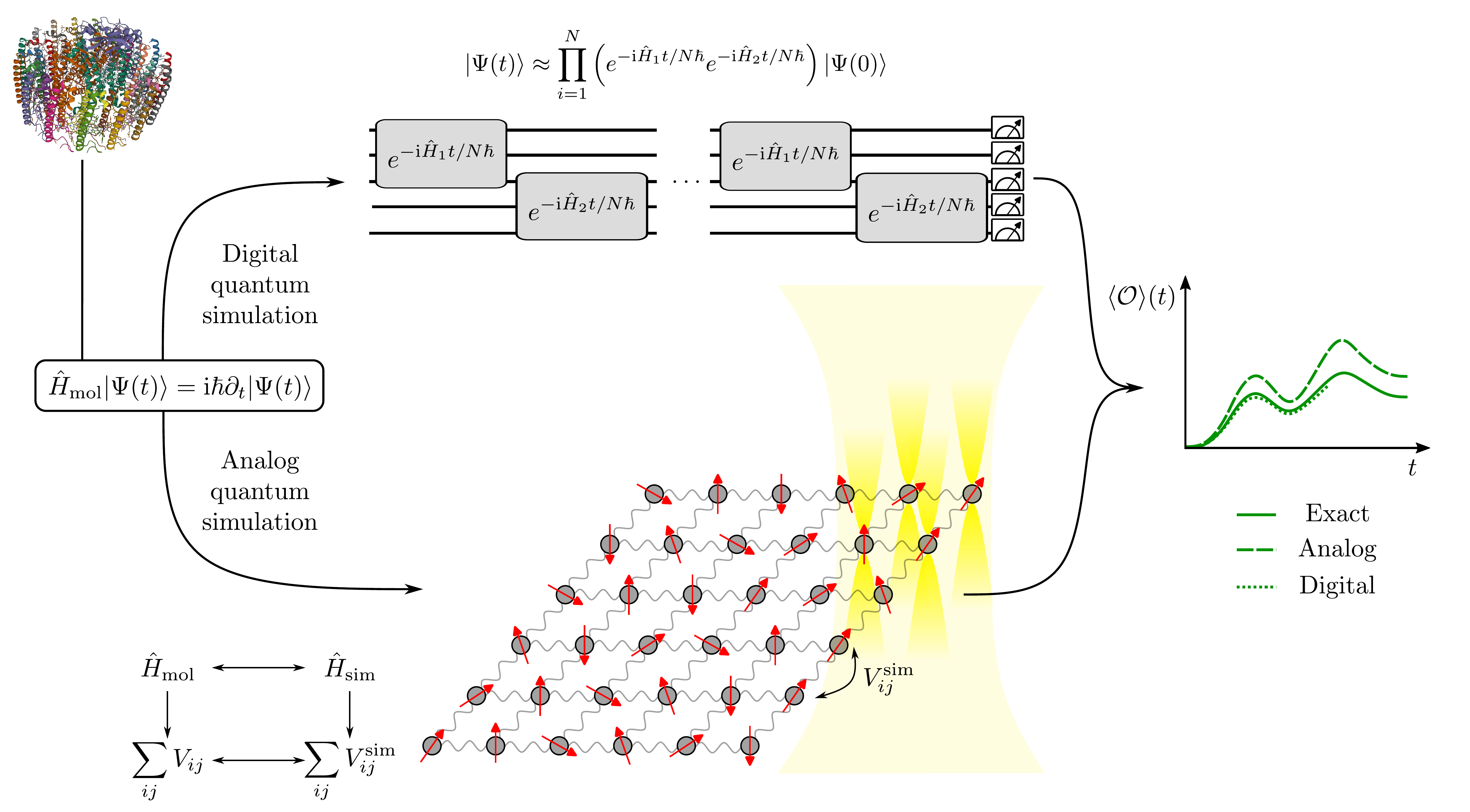}
  \caption{Graphical representation of the steps involved in the simulation of a molecular process with digital and analog quantum simulation.
  For a given biochemical target (such as the LHC-II complex in the upper left part of the figure), one first chooses
  the Hamiltonian $\hat{H}_\text{mol}$ that describes the process at hand.
  For LHC-II, this could be the excitonic Hamiltonian describing the excitation energy transfer process.
  Assume we aim at solving the time-dependent Schr\"{o}dinger equation (box in the left part of the figure) for a given initial wave function $\vert \Psi(0) \rangle$.
  In digital quantum simulations (upper part of the figure), one first maps the target quantum system to a set of $N_\text{q}$ qubits (in the figure $N_\text{q}=5$, and maps $\hat{H}_\text{mol}$ onto a corresponding qubit Hamiltonian.
  A sequence of gates
  is then defined to solve the time-dependent Schr\"{o}dinger equation.
  In the figure, we selected the specific case of the Trotter algorithm for simulating quantum dynamics (see Eq.~(\ref{eq:Trotter})).
  In analog quantum simulations (lower part of the figure), the quantum system is first mapped onto a set of interacting quantum entities (such as, for instance, atoms trapped in arrays of optical tweezers).
  The interaction between these entities is then tuned (for instance, with a laser, as represented in the figure) such that the Hamiltonian describing these interactions ($\hat{H}_\text{sim}$) mimics the original Hamiltonian $\hat{H}_\text{mol}$.
  The quantum dynamics is then simulated \emph{directly} on the resulting simulator, without the need of designing any algorithm for solving the time-dependent Schr\"{o}dinger equation.
  The result of a simulation is a time-dependent property $\langle \mathcal{O}\rangle (t)$, as represented in the right part of the figure.
  The analog quantum simulation does not admit rigorous approximation bounds yet aims to give an accurate qualitatively reliable simulation for a long propagation time.
  The digital simulation, by contrast, will track the property with high accuracy, yet only for a shorter period of time due to the stringent hardware requirements of fault-tolerant quantum computation.}
  \label{fig:DigitalAnalog}
\end{figure}

\emph{Digital quantum simulators} are, by contrast, based on the gate-based model of quantum computation in direct analogy to the gate-based model of standard (classical) computation (see the upper part of Figure~\ref{fig:DigitalAnalog}).
Here, a quantum algorithm is cast as a sequence of quantum gates, \textit{i.e.}, operations acting on quantum bits (qubits).
Qubits are the analog of the bits in a standard computer, but can, in addition to the states 0 and 1, also assume a quantum superposition of both, in the same way that an arrow cannot only point up or down but also in any other direction.
In a physics context, qubits are also known as two-level systems and can often be viewed as a spin-$\frac{1}{2}$ system.
Importantly, the composition rule of quantum particles, and hence also of quantum bits, posits that the state of $n$ qubits requires the description with $2^n$ probability amplitudes (i.e., expansion coefficients; see Figure 3)
describing the precise superposition of the strings of $n$ bits (thereby making up the `state' or `wave function' of the $n$ qubits).
Simulating or even just storing the amplitudes on a standard computer, therefore, requires more than $2^n$ bits and therefore an exponential overhead (the curse of dimensionality eluded to earlier).

Whereas the storage of classical bits is stable, as the `0' or `1' is copied many times over, for instance, in the many spins of a magnetic patch storing a bit on a hard drive, the situation is markedly different in quantum computation.
Quantum information cannot be copied and is therefore much more fragile and affected by quantum noise, the previously mentioned decoherence.
Consequently, storage and gate operations on quantum bits are prone to errors that need to be corrected.
Note that quantum error correction, first put forward by Shor in 1996,\cite{Shor1995_QuantumErrorCorrection} alone is not sufficient, because the quantum circuits carrying out the error correction may themselves introduce new errors.
Luckily, techniques have been developed that replace each noisy \emph{physical qubit} and physical gate in a circuit with an encoded \emph{logical qubit} and logical gate, thereby allowing an entirely noise-free quantum computation to be simulated on noisy quantum hardware (given that the noise is below a threshold -- the `threshold theorem')\cite{AB99}.
In addition to the quantum hardware and quantum algorithms, quantum computers therefore also require an important error correction layer in between.
That fault-tolerant quantum computation is possible at all is mathematically highly non-trivial, but allows us to imagine that digital quantum simulation is possible in the future, despite decoherence possibly staying around indefinitely.

There is a price to be paid for the fault-tolerant hardware
in terms of the overhead required.
Whereas recently, it has been shown that only a constant overhead is necessary (i.e. the number of logical qubits is proportional to the number of physical ones)\cite{fawzi-constant-2021} and one may therefore generally see no fundamental obstacle to building a perfect digital quantum simulator in the future, the number of physical qubits required for the first viable algorithms to run is extremely large and therefore not expected to be reached in the near future, but only in the medium or long term.
We note that one may hope for novel hardware solutions that lead to a lower overhead due to stronger protection of the native qubits to noise, e.g., through topological protection.\cite{Kitaev1997_ErrorCorrection,Kitaev2003_Anyons}

Therefore, research on quantum algorithms splits, on the one hand, into algorithms for fault-tolerant quantum computers, which we may think of as a future perfect quantum computer, and, on the other hand, into algorithms for near-term or noisy-intermediate stage quantum (NISQ)\cite{Preskill2018_NISQ} devices of dozens to thousands of physical qubits, that are characterized by the absence of complete fault-tolerance.

Naturally, the algorithm design focuses on rigorous runtime bounds for the fault-tolerant machines and heuristic arguments for NISQ devices.
Common to problems from both chemistry and molecular biology is the necessity of mapping the representation of electrons onto the qubit structure.
This mapping is often done with the Jordan-Wigner\cite{Jordan1928_JW-Transformation} or the Bravyi-Kitaev transform,\cite{Bravyi2002_Transform,Love2012_BK-Transform}, but further algorithms-tailored methods are expected to be developed in the future.
Common to all simulation algorithm designs is the drive to the most favorable scaling of simulation time with the size of the simulated system in order to make actual quantum computations feasible.

An important class of algorithms for NISQ hardware are so-called hybrid quantum-classical algorithms that partition the computational workload between a classical and a quantum computer.\cite{McClean2016_VQE,Cerezo2022_VQE}
The most prominent example of this class of algorithms
relevant for electronic structure calculations is the variational quantum eigensolver (VQE), which presents a variational (and thus heuristic) algorithm to upper bounding the ground state energy problem.\cite{Peruzzo2014_VQE,Kandala2017_VQE-Experiment,Tilly2022_VQE-Review}
In a nutshell, a given wave function ansatz is represented with a parametrized quantum circuit (i.e. a quantum circuit,
where the individual gates are dependent on parameters), and its energy is estimated with a term-by-term measurement.

As we have discussed, there exist wave function ans\"atze inspired by the ones described in the previous section for which the cost of measurement would grow exponentially on classical hardware and polynomially on quantum hardware, therefore allowing to circumvent the exponential overhead.
However, VQE yields the exact solution to the full-CI problem only if the final wave function can be represented with that ansatz.
The parameters of the wave function ansatz are then optimized in an outer loop on a classical computer taking the quantum energy measurements as input in order to obtain a variational upper bound to the true ground state energy.

Other examples of hybrid classical-quantum algorithms are the variational quantum time evolution for quantum dynamics simulations\cite{Benedetti2020_VariationalQuantumTD,Barison2021_EfficientQuantumTD} and the variational imaginary-time evolution,\cite{McArdle2019_QuantumImaginaryTimeEvolution} which represents an alternative to VQE for ground state energy optimizations.

On a fault-tolerant computer, one may directly measure the full-CI energy by first preparing a wave function ansatz on the quantum hardware (as in VQE) and then applying the Quantum Phase Estimation algorithm, which projects a prepared state into an energy eigenstate of the Hamiltonian.\cite{lloyd-1996,AspuruGuzik2005_QPE-Original,AspuruGuzik2008_QPE-MolecularEnergies,Veis2010_QuantumComputing-CH2,Veis2012_RelativisticQuantum,Veis2016_QuantumComputing-PreBO}
The ground state will be found with high probability after $O(\frac{1}{\kappa})$ repetitions, where $\kappa$
is the overlap between the prepared state and the exact ground-state wave function.
Note, however, that the ground state energy problem is known to be QMA-hard
and it might, therefore, be expected that no general exponential speedup will be obtained, rendering this again a heuristic algorithm.
In practice, however, for those electronic structure problems that are dominated by dynamic correlation (see above), the Hartree-Fock state will be a reliable initial state with sufficient overlap with the final state, as this has been the key insight into the development of single-reference approaches such as coupled cluster for decades.
In the presence of strong static correlation, the Hartree-Fock state may not have a sufficiently large overlap with the final wave function (see Ref.~\citenum{Chan2022_ExponentialAdvantage}). However, in these cases, an amazing qualitative and quantitative understanding of good candidates for functions (such as the FCI-type wave functions optimized by DMRG or FCIQMC; see above) has been assembled that approximate such strongly correlated systems well so that there is hope that the overlap $\kappa$ indeed can be made substantial.

For Quantum Phase Estimation, one is required to perform a time evolution with a Hamiltonian that describes the (bio)molecule under study with parameters that have been calculated as integrals in a so-called four-index transformation from the atomic orbital basis set to the molecular orbital set.
Since the Hamiltonian consists of a sum of terms, the basic idea is to use a Lie-Trotter formula in order to be able to decompose the generally complicated time evolution into gates
that only touch two particles at a time. For a Hamiltonian with two terms $H=H_1+H_2$ time-evolved for time $t$,
this yields the following approximation

\begin{equation}
  e^{-\mathrm{i}tH/\hbar} = e^{-\mathrm{i}tH_1/\hbar - \mathrm{i}tH_2/\hbar} \approx
    \underbrace{\left( e^{-\mathrm{i}H_1t/N\hbar} e^{-\mathrm{i}H_2t/N\hbar} \right)
                \left( e^{-\mathrm{i}H_1t/N\hbar} e^{-\mathrm{i}H_2t/N\hbar} \right) \cdots
                \left( e^{-\mathrm{i}H_1t/N\hbar} e^{-\mathrm{i}H_2t/N\hbar} \right)}_{N \text{terms}} \, ,
  \label{eq:Trotter}
\end{equation}

whose quality improves for increasing $N$. Note that it is a sequence of shorter propagations each time only involving one of the terms in the Hamiltonian.
By now, significant improvements on this initial decomposition idea have been devised and an excellent, indeed asymptotically optimal, understanding has been obtained\cite{Chuang2017_QuantumSignalProcessing,childs-2019}.
Yet, significant further development is needed, especially in regard to concrete and tailored instances of simulation problems.

Finally, we emphasize that whether analog or digital, fault-tolerant or NISQ, it is important to realize that asymptotic runtime bounds will only be a guide to the performance
and practical resource estimates will have to be obtained.
These bounds will additionally depend strongly on the available hardware, e.g. on the number, the connectivity, and the control of the qubits, on the gate set, and the type of noise.
To make practical use of quantum computation for molecular biology (or molecular simulations in general), all of these aspects of the available hardware need to be combined with an in-depth knowledge of the molecular system(s) at hand and, therefore, might require highly specialized knowledge and expertise ranging from molecular biology to the understanding of the state of the quantum hardware to eventually the quantum algorithm design.

In the remaining parts of the manuscript, we will detail the prospects of quantum computation in several distinct areas of molecular biology.

\section{Local quantum effects in biochemical processes}
\label{sec:LocalQuantumEffects}

We will now turn our attention to (local) quantum mechanical phenomena in molecular biology, which mostly relate to chemical reactions of biomolecules in the most general sense (not only covering stoichiometric or enzymatic reactions, but any kind of chemistry of biomolecules, even involving solvation and association processes).
Chemical processes have been recognized as a potential high-impact application of future fault-tolerant quantum computers\cite{Reiher2017_PNAS,Reiher2017_CEN}.
Accordingly, chemical reactions in biomolecular systems are an obvious target for quantum computations with significant advantages under certain conditions.
In fact, such a scenario has been first elaborated in detail for the mode of action of the enzyme nitrogenase\cite{Reiher2017_PNAS} and was further expanded on for other catalytic processes with improved quantum algorithms.\cite{vonBurg2021_Catalysis-QuantumComputing}

The key feature exploited in this context is the spatial locality of chemical reactions; i.e., they involve comparatively few atoms that are involved in dissociation and association (bond-breaking and bond-formation) processes.
The huge rest of the biomolecular machinery is treated as a spectator environment in a given chemical reaction step.
This environment can be efficiently described in a classical embedding scheme, an idea that won the Nobel Prize in Chemistry in 2013.\cite{Warshel2014_NobelLecture}
As a result, the complex environment can be efficiently modeled and only the local quantum description of the reaction event remains the major challenge: the biochemical problem has been reformulated as a purely chemical problem, and all developed solution approaches in the chemical context automatically transfer to the biochemical application.

In reaction chemistry, these processes are all related to changing the connectivity of atoms in molecules, i.e., to changing  covalent bonding in some way.
Examples are enzymatic reactions, but also any kind of chemical reaction that carries biological significance.
Hence, it starts already with the creation of the first biomolecules (abiogenesis) and the origin of life. Accordingly, quantum computation promises to advance not only the accuracy of quantum predictions for enzyme kinetics, especially for strong correlation cases as they
are found in metalloenzymes with $3d$ metal-atom active sites such as copper and iron-sulfur enzymes (including electron and proton transfer processes), but also for stoichiometric reactions and biochemical pathways in living cells.

Naturally, any chemical reaction network of relevance to molecular biology can benefit from quantum computation.
Intellectually appealing examples may even be taken from prebiotic chemistry (see, for example, the formose reaction network\cite{AspuruGuzik2014_FormoseReaction,Simm2017_ContextDriven}), where (traditional and quantum) simulations may provide insights into prebiotic potentialities that have not been scrutinized by experiment yet.
Quantum computation might even deliver accurate electronic energy estimates for reactive biomolecular precursors on surfaces that act as catalysts for the production of the first (functional) biomolecules (cf. the Waechtershaeuser hypothesis\cite{Wacthershauser1988_IronSulfur,Wacthershauser1992_IronSulfur}).

Even more relevant would be the application of quantum computing to simulate reactions that are difficult to reproduce in a laboratory setting.
An example are astrochemical reaction networks emerging under the extreme conditions of extraterrestrial environments.
In the context of molecule formation in interstellar space, often the lack of an environment (producing conditions far away from thermal equilibrium) will require high-accuracy rovibronic quantum-state resolved formation theories and corresponding calculations.
The high accuracy needed to reliably predict the outcome of these reactions, combined with the relatively small size of the intermediates entering astrochemical reaction networks, makes astrochemistry\cite{Barone2015_AstrochemistryReview,Oberg2016_AstrochemisryReview,Lee2020_Astrochemistry} an appealing target for quantum computing.
In the context of astrobiology, biomolecules (such as amino acids) on surfaces of interstellar objects (exoplanets or asteroids) subjected to the light of stars may well fall into the category of approaches applicable to pre-biotic scenarios on Earth.\cite{Hand2005_Astrobiology}
Moreover, computer simulation can even adjust the conditions to account for the macroscopic constraints such as increased temperature and pressure.
Hence, within this context, potential routes leading to the emergence of life in outer space could also be scrutinized based on quantum computing simulations.

In the physical description, chemical processes are mapped to a path across the Born-Oppenheimer PES (cf. Figure~\ref{fig:IntroductoryFigure})
so that the calculation of the electronic energy as a solution of the electronic Schr\"odinger equation is the key first step (cf. Figure~\ref{fig:BornOppenheimer}).
Accordingly, the accurate quantum mechanical calculation of the electronic energy for a given arrangement of the atomic nuclei in the biomolecules under consideration suffers from the curse of dimensionality through the large number of orbitals required for accurate results.

In the past decades, a zoo of practical solution methods has been devised for traditional computing hardware (see Section~\ref{sec:TraditionalApproaches}).
However, every method strikes a compromise between feasibility and accuracy: the more accurate the energies (and, in particular, the energy differences) need to be, the smaller the systems (in terms of the number of atoms or electrons) are for which an accurate method is feasible on some traditional computer hardware.
Still, these traditional approaches have reached a remarkable degree of sophistication, accuracy, usefulness, and acceptance.\cite{Pople1999_NobelLecture,Kohn1999_NobelLecture,Dykstra2005_Book}
Detailed resource estimates\cite{Reiher2017_PNAS,Liu2022_QuantumComputingReview} showed that algorithms based on quantum phase estimation are likely to compete and beat the best traditional approaches for cases of strong electron correlation as they occur in active sites of metalloenzymes (and in electronically excited states; see below).
Their main advantage though is the explicit control over the error bar on the result (provided that the initial state prepared features sufficient overlap with the final state; see discussion in the previous section), which is not available from any traditional approach and also not from the currently most favored algorithm for NISQ devices, namely, VQE with a unitary coupled cluster ansatz because of limitations on the underlying parametrization.

\section{Global quantum effects in functional biomolecules}
\label{sec:GlobalQuantum}

\begin{figure}[htbp!]
  \centering
  \includegraphics[width=.85\textwidth]{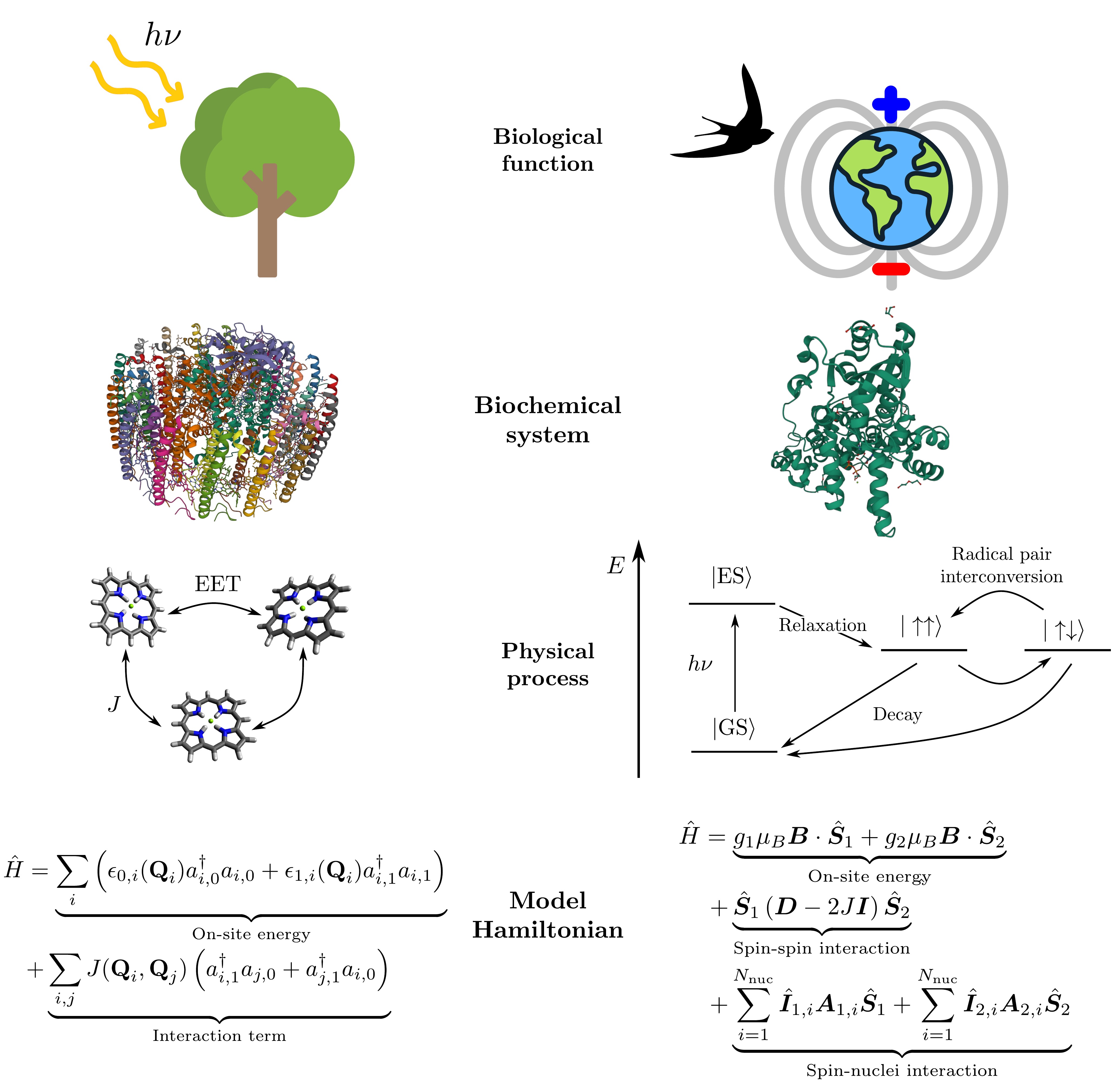}
  \caption{Key steps involved in the modelling of photosynthesis (left part) and magnetoreception (right part; "Biological function").
  First, the biomolecule that tunes the target biological function is identified ("Biochemical system"):
  a light-harvesting complex for photosynthesis and a cryptochrome protein for magnetoreception.
  Second, the "Physical process" linked to the biological function of the molecules is identified.
  For photosynthesis, this is the excitation energy transfer (EET) that enables to funnel the energy captured by the chromophores through light absorption to the reaction center.
  For magnetoreception, the key physical process is the generation of the radical pair in the FAD-tryptophan complex inside the cryptochrome.
  Finally, one defines a ``Model Hamiltonian'' that describes the quantum process which tunes the biochemical function.
  For the EET in a light-harvesting complex, the shown excitonic Hamiltonian can describe the chromophores as two-level quantum systems (with energies $\epsilon_{0,i}$ and $\epsilon_{1,i}$ for the $i$-th chromophore, both of which depend on the chromophore nuclear coordinates $\bm{Q}_i$), that interact \textit{via} a nearest-neighbor potential.
  For the radical pair dynamics, each electron of the radical pair is described by two spin operators, $\hat{\bm{S}}_1$ and $\hat{\bm{S}}_2$ that interact with the Earth magnetic field $\bm{B}$, among themselves \textit{via} a dipolar spin-spin interaction ($\bm{D}$ is the dipolar interaction tensor, whereas $\bm{I}$ is the unit matrix), and with the nuclear spin (referred to as $\hat{\bm{I}}_k$ for the $k$-th nucleus).
  Note that a system as large as the cryptochrome contains several dozens of nuclei with non-zero spin.}
  \label{fig:Excitonic}
\end{figure}

We highlighted in the previous section that a key feature of quantum effects associated with the rearrangement of nuclei and electrons that drive biochemical reactions is their locality in space -- \textit{i.e.} they are relevant only for the reacting nuclei and their electrons (and for a comparatively tiny fraction of their environment).
Other biomolecular processes are, instead, influenced by quantum effects acting on a larger spatial scale. For instance, large amplitude motions of solvated proteins are, at their very heart, governed by dynamic electron correlation effects.
While these can often be well modelled by classical surrogate potentials, this is not always the case.
Two prominent examples are \emph{light harvesting} and \emph{magnetoreception} (see Figure~\ref{fig:Excitonic}).
Light harvesting represents the initial step of photosynthesis, \textit{i.e.} the process exploited by plants to convert light into chemical energy.
Photosynthesis proceeds \textit{via} a chain of complex chemical processes, the first of which is light harvesting, which corresponds to the absorption of visible light by so-called pigment-protein complexes.\cite{Walters1996_LightHarvestingReview, Fleming2009_LightHarvesting-Review,Scholes2011_LightHarvestingReview}
Magnetoreception\cite{Schulten2000_Magnetoreception} is the process that enables birds to sense weak magnetic fields and, in particular, the Earth magnetic field that can be used as a guide then.
It has been argued that quantum effects play a key role in both phenomena.

In contrast to the chemical reactions described in the previous section, quantum effects span, in this case, the whole molecule.
These phenomena are the subject of study of so-called \emph{quantum biology}.\cite{Nori2013_QuantumBiology,Plenio2013_QuantumBiology,Zigmantas2020_QuantumBiology}
We prefer not to use this term because the effects described in Section~\ref{sec:LocalQuantumEffects} have also a quantum nature, but are usually not referred to in the realm of quantum biology.

Pigment-protein complexes encapsulate several chromophores, usually porphyrin derivatives.
The light is first absorbed by the chromophores.
The resulting electronic excitation is then transferred, \textit{via} a chain of excitation energy transfer (EET) processes, to the reactive center that converts it into chemical energy.\cite{Mennucci2017_LightHarvesting-Review,Scholes2017_Review}
The overall efficiency of photosynthesis depends critically on whether the electromagnetic energy of the absorbed light can be transferred to the reaction center without losses.
Multidimensional electronic spectroscopy measurements of light-harvesting systems have revealed the presence of quantum coherences (\textit{i.e.}, that the electronic wave function is a coherent superposition of excited states centered on each chromophore) in the early phase of the EET process.\cite{Engel2007_QuantumCoherences,Engel2010_QuantumCoherences,Collini2010_QuantumCoherence}
This observation, combined with the fact that quantum mechanical models predict a higher efficiency compared to semiclassical calculations, has led to the hypothesis that light harvesting is driven by quantum effects.
Although it was later observed that the lifetime of electronic coherences may not be long enough to have an effect on the EET rate,\cite{Miller2017_RelevanceQuantumEffects,Thyrhaug2018_DifferentCoherences} it has been argued that molecular vibrations, through their coupling with the electronic excitation (usually referred to as excitonic coupling), may enhance the coherence lifetime by quantum coherences.\cite{Chin2013_NonEquilibriumVibrations,OlayaCastro2014_Vibronic,Scholes2016_Vibronic}

Similar observations have been made for magnetoreception.
According to the currently most favored hypothesis, magnetoreception relies on the generation of a radical pair in so-called cryptochromes.\cite{Schulten1999_Review,Schulten2001_Review,Schulten2010_Review}.
The latter are proteins located, in particular, in the bird's eyes, where they can absorb visible light.
Light absorption triggers a chain of electron transfer reactions that generates two unpaired electrons (the radical pair) localized on a flavin adenine dinucleotide and a tryptophan residue of the protein.
The spin state of the radical pair can be either singlet or triplet, and an external magnetic field can affect the relative amount of these two states.
Since the rate at which the radical pair decays to the electronic ground state \textit{via} spin recombination depends on the spin state, a magnetic field ultimately influences the reaction outcome.
This magnetic field sensitivity is believed to be at the origin of magnetoreception.\cite{Hore2009_Magnetoreception,Hore2016_Magnetoreception}
Semiclassical simulations predict a radical pair lifetime that is much shorter than that obtained with quantum-mechanical models.\cite{Fay2020_QuantumMagnetoreception}
As for light harvesting, this has been taken as an indication that quantum mechanical effects play a key role in magnetoreception.

For both phenomena, quantum mechanical effects affect the interaction between molecules -- the chromophores, for light-harvesting, and the unpaired electron, for magnetoreception.
Simulating quantum effects happening on such large spatial scale is much harder than for chemical reactions.
The electronic wave functions of these systems depend on thousands of electrons, and this makes applying the theory described in Section~\ref{sec:TheoreticalFoundation} extremely challenging for both classical and quantum computations.
As a way out, effective Hamiltonians (usually referred to as \emph{excitonic Hamiltonians}) have been designed that largely reduce the problem complexity but still encode the key quantum mechanical effects that drive these biological phenomena.

In these Hamiltonians, the chromophores are approximated as quasiparticles, the excitons, with two quantum levels each, one corresponding to the electronic ground state and the other one to the excited state.
The inter-chromophore interaction is then approximated with effective potential energy terms that account for the excitation energy transfer phenomenon.
Condensing all the electrons of a chromophore into a two-level quantum system drastically reduces the Hamiltonian dimensionality and, consequently, the complexity of solving the corresponding quantum equations, but at a huge loss in terms of detail and resolution.
Excitonic Hamiltonians can be further extended to include also the nuclear degrees of freedom, which leads to so-called (phenomenological) vibronic Hamiltonians.

Even within the excitonic approximation, solving exactly the Schr\"odinger equation for a faithful Hamiltonians describing a pigment-protein complex or a radical pair is extremely hard with classical computing algorithms.
In the case of light harvesting, a pigment-protein complex may be composed of more than 10 chromophores, each one with several dozens of degrees of freedom.
For magnetoreception, the radical pair interacts with the nuclear spins.
The number of nuclei with non-zero spin may be large, which makes simulating the radical pair dynamics computationally hard.
In both cases, this results in many-body Hamiltonians depending on more than 100 particles, which is out of the reach of the classical simulation strategies described in Section~\ref{sec:TheoreticalFoundation}.
For this reason, the claim that global quantum effects are relevant in photosynthesis and magnetoreception relies on simulations of simplified problems that further approximate the biochemical process.
For instance, quantum mechanical simulations of magnetoreception have been limited to radical pairs that are much smaller and simpler than the cryptochromes.\cite{Plenio2013_ChemicalCompass,Manolopoulos2016_Magnetoreception,Manolopoulos2020_Magnetoreception}
Quantum computing might have the potential to overcome these limits, provided that either a sufficiently large number of logical qubits can be provided in huge fault-tolerant quantum computers or a large analog quantum simulator can be built that can represent Hamiltonians with so many quasiparticles.

In fact, both the efficiency of the EET process
in pigment-protein complexes and the lifetime of the radical pair can be predicted by solving the time-dependent Schr\"odinger equation for the corresponding Hamiltonian.
As we discussed in Section~(\ref{sec:quantum}), this is one of the few examples of a problem for which theoretical guarantees of an exponential speedup can be derived.
In practice, the major challenge will be ensuring quantum hardware stability over the whole propagation time.
As we display Figure~\ref{fig:DigitalAnalog}, achieving such stability is particularly hard in digital quantum computing.
Conversely, quantum simulators may be better suited for this class of simulations.
As a matter of fact, some pilot experimental realizations of quantum simulators for light-harvesting systems have been proposed in recent years,\cite{Wang2018_QuantumSimulation-LHC,Potocnik2018_QuantumSimulation-LHC} although they are limited to very simple excitonic Hamiltonians.

\section{Quantum computing for structural biology}
\label{sec:QuantumForClassical}

Processes driven by quantum effects induced either by the breaking or forming of covalent bonds (see Section~\ref{sec:LocalQuantumEffects}) or by light excitation (see Section~\ref{sec:GlobalQuantum}) represent only a very small portion of the realm of biological processes.
In fact, we already highlighted in the introduction that molecular biology is governed by reversible, non-covalent interactions between biomolecules.
Although these interactions can be described with the quantum approaches presented in Section~\ref{sec:quantum}, such a route is both impractical and superfluous.
It is impractical because, due to the curse of dimensionality, full atomistic quantum mechanical simulations of large biomolecules will most likely remain out of reach for any simulation method, both with classical and quantum computation.
Furthermore, it is superfluous, because weak interactions are effectively approximated by classical surrogate potentials, i.e., by the so-called \emph{force fields}.\cite{Scott1999_Gromos,Wang2000_Amber,Ponder2003_MartiniForceField,HeadGordon2018_ForceField-Review}
A force field approximates the PES of a molecule through a classical parametric interaction potential between its atoms, as shown in Figure~\ref{fig:IntroductoryFigure}.
The interaction energy between atoms that are linked by a covalent bond is expressed as a power series in terms of bond lengths $r_i$ and angles $\theta_j$, and as a Fourier series in terms of the dihedral angles $\gamma_m$ (see Figure \ref{fig:IntroductoryFigure}).
The non-covalent dispersion-repulsion interaction between non-bonded atoms is, instead, expressed as a sum of pairwise Lennard-Jones-type interactions.
The parameters entering the force field (\text{i.e.}, $k_i^\text{(b)}$, $k_j^\text{(a)}$, $k_m^\text{(d)}$, $C_6^{ij}$, and $C_{12}^{ij}$ in Figure~\ref{fig:IntroductoryFigure}) are obtained by least-squares fitting either to results of quantum chemical calculations or to experimental data.
The force field then defines a potential that can be used in classical atomistic simulations based on the Newton equation of motion of the atoms.

\begin{figure}[htbp!]
  \centering
  \includegraphics[width=\textwidth]{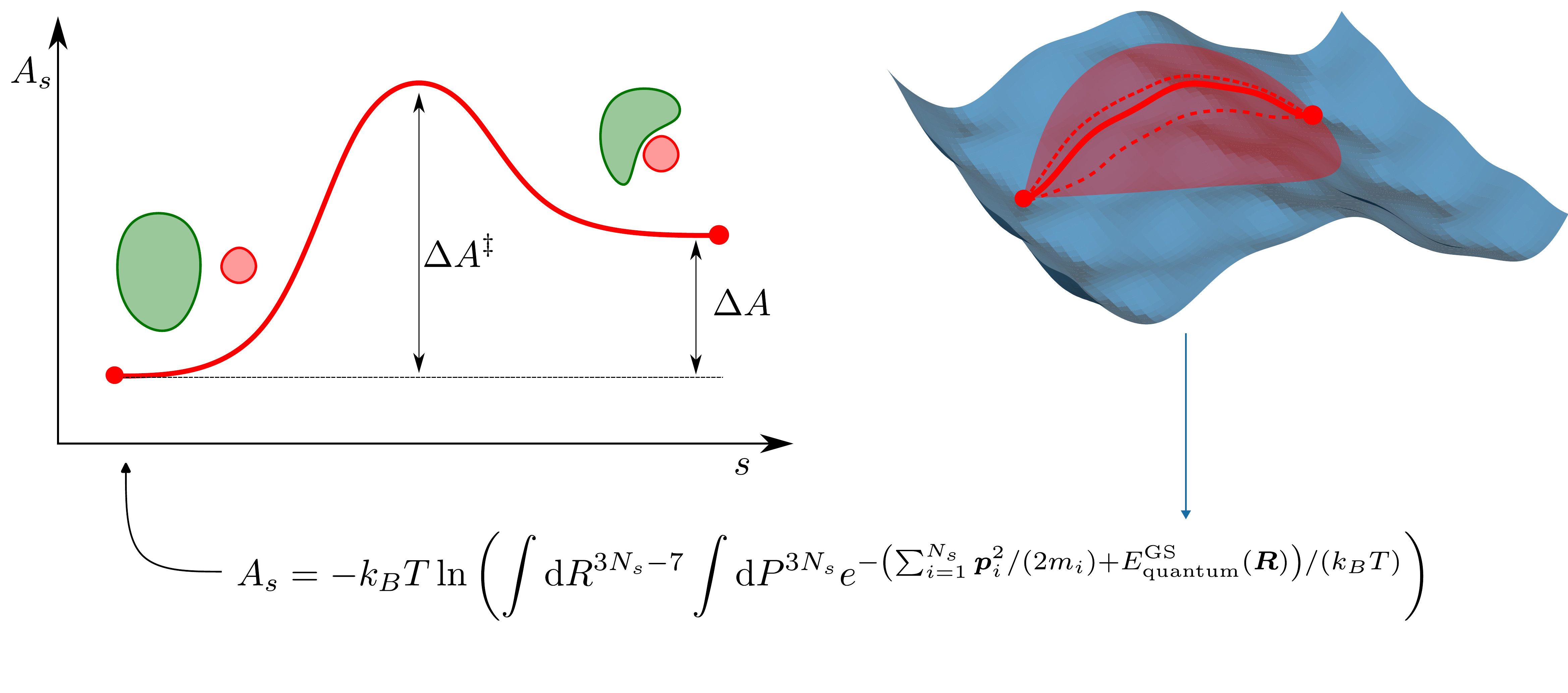}
  \caption{Graphical representation of the steps needed to calculate the potential of mean force for a biochemical process.
  The process coordinate $s$ describing, for instance, the binding of a small molecule (represented as a red circle) to a protein (represented as a green object)
  may be some distance between these two molecular systems.
  The contribution of all the $3N_s-7$ remaining coordinates (where $N_s$ is the number of atoms in the molecular system comprised of the red and the green molecule) is averaged out by Boltzmann weighting in a canonical ($N$, $V$, $T$) ensemble. For the single molecular system, this is represented by the integral reported in the lower part of the figure, yielding the classical partition function for the molecular system.
  The potential energy entering the Boltzmann factor is the electronic energy (i.e., the Born-Oppenheimer hypersurface) or its approximation by a classical surrogate potential defined in Figure~\ref{fig:IntroductoryFigure}.
  The logarithm of the integral of the classical partition function is proportional to the (molecular) potential of mean force $A_s$. The Helmholtz free energy $A$, which depends on the system volume $V$, its temperature $T$, and the macroscopic number of molecular systems $N$, is obtained by integrating over each $s$ in every molecular system in the macroscopic system.
  The free-energy difference between reactants and products $\Delta A$ determines wheter a given process is thermodynamically favorable, while the transition-state free-energy difference $\Delta A^{\ddagger}$ determines the corresponding rate constant.}
  \label{fig:PotentialOfMeanForces}
\end{figure}

At this point, it is important to highlight that any biological process cannot be considered an isolated process.
In fact, it happens at about room temperature and involves a continuous energy exchange with the environment (for instance, the cell in which a protein is located).
A biomolecule is then not to be represented by a single molecular structure, but rather as a thermal ensemble of many configurations, each one weighted by the appropriate thermodynamic weight.
The key quantity is, therefore, not the energy of the microstate of a single configuration, but the free energy calculated from the partition function of all microstates.
The partition function also gives access to all thermodynamic properties.
The PES is not the appropriate quantity to interpret the reactivity within such a thermodynamic ensemble as the driving force of a reaction is, in fact, the free energy (see Figure~\ref{fig:PotentialOfMeanForces}).
Typically, the \textit{potential of mean forces}\cite{Kirkwood1935_PotentialOfMeanForces} is then considered, which generalizes the concept of PES to thermal systems.
For a given reaction profile (e.g., one of the paths of Figure~\ref{fig:IntroductoryFigure}), the potential of mean forces is calculated, at each point of the PES, by averaging out all degrees of freedom of the molecule but the reactive one.
Despite this conceptual difference, the electronic energies remain the cornerstone also for calculating the potential of mean forces because they are a key contribution to the partition function and, therefore, to any thermodynamic quantity.
One may wonder whether quantum computing may be applied to speed-up also this step.
While algorithms to calculate the partition function of quantum systems on quantum hardware have been proposed,\cite{Motta2020_ThermalQuantum} which is not surprising in view of the \textit{quantum simulates quantum} principle introduced above, extensions of these schemes to compute classical thermal averages are still scarce.
Quantum computing realizations of the Monte-Carlo algorithm exist,\cite{Temme2011_QuantumMetropolisSampling} and so it may be argued that they can be applied to calculate thermal averages \textit{via} sampling techniques.\cite{Frenkel2002_UnderstandingMolecularSimulation,Micheletti2021_Polymer-QuantumComputing,Ghamari2022_QuantumSampling}
However, these algorithms assume that the distribution function can be encoded as the probability distribution of a many-body wave function (see Section~\ref{sec:TheoreticalFoundation}).
Hence, applying them to calculate classical thermodynamic averages requires mapping a classical probability distribution onto the qubit states, a task that is hard itself, as we will discuss in the following section.

Once an appropriate force field is available, simulating the classical dynamics of a biomolecule represents the cornerstone of several types of biological simulations.
The most prominent example is the protein folding problem, \textit{i.e.} the task of predicting the three-dimensional structure of a protein based on its amino acid sequence.
The protein folding problem can be solved by simulating the dynamics of a protein under thermal equilibrium conditions.
For sufficiently long propagation times, the protein will evolve into its most stable conformation, thereby yielding the solution to the protein folding problem.
The three-dimensional structure of a protein can be reliably predicted with force fields because it is mostly determined by the non-covalent interactions between the amino acid residues of the protein.
Not being affected by the curse of dimensionality, classical simulations can target much larger molecules compared to quantum mechanical simulations so that folding simulations of small fast-folding proteins are possible.\cite{Kollman1998_MD-Folding,Snow2002_ProteinFoldingSimulations,Shaw2011_Folding}

Since classical simulations are not impacted by the curse of dimensionality, one may argue whether quantum computing can be beneficial also to this field.
The first, natural combination of these two fields is to parametrize a force field based on energies obtained from quantum computations.
It would be possible, for instance, to parametrize or refine a force field that describes molecular interactions in proteins based on exact quantum chemical simulations on small polypeptides.
A small polypeptide may contain a few dozen atoms and a few hundred electrons and is, therefore, too large to be targeted by classical quantum chemical calculations.
Such a molecular size may be, in the near future, within the application range of quantum computing, in which case such a route would provide very accurate data for parametrizing force fields (provided that the quantum machine can represent the discretized problem well).
This idea has been recently explored in Refs.~\citenum{Malone2021} and \citenum{Kirsopp2022_ProteinLiganQuantum-DMET}, where VQE-based methods delivered data for parametrizing classical force fields.
In the future, the very same problems can be explored with algorithms for fault-tolerant quantum computers.

Quantum computing can also be applied to classical simulations that rely on a given force field.
For instance, protein folding can be rephrased as an optimization problem.
The stable structure of a protein corresponds, in fact, to the structure yielding the lowest free energy.
However, this optimization problem is classically complex (NP-hard) to solve.\cite{Fraenkel1993_FoldingNPHard}
For this reason, the folding problem can be solved based on this idea only for simplified lattice models that drastically reduce the conformational freedom of a molecule.\cite{Levitt1992_Folding}

Note that protein folding is not the only molecular biology problem that can be rephrased as an optimization problem.
Another example is molecular docking, \textit{i.e.} the prediction of the most stable binding configuration of a ligand to a given target, which represents the first step of \textit{in silico} drug design.
However, also in this case, the optimal configuration may be obtained by minimizing a classical force field describing the interaction between the ligand and the target.
As for the protein folding, the resulting optimization problem is classically hard.\cite{Halperin2002_Docking,Dias2008_Docking}

Since optimization problems may be one of the first and primary application fields of quantum computing, the question arises of whether quantum computing may boost the efficiency of optimization-based molecular biology algorithms.
Whereas polynomial-time quantum algorithms are not expected for NP-hard problems,\cite{Vazirani1997_QuantumComputingHardness} like general optimization problems, more modest speedups, still significant in practice, might be achievable.
An example is the quadratic speedup that can be achieved based on an exhaustive search for the optimal solution with Grover's search algorithm\cite{Grover1997_SearchAlgorithm}
(leaving aside questions of efficient data transfer and storage in the quantum machine; see below).

Moreover, quantum optimization algorithms for which a gain in performance cannot be guaranteed, but has been observed heuristically for small test cases, have been designed.
In the latter category, the quantum approximate optimization algorithm (QAOA)\cite{Fahri2014_QAOA} is noteworthy and is currently under in-depth investigation.\cite{Lloyd2018_QAOA-Universality,Harrow2016_QAOA,Lukin2020_QAOA,Harrigan2021_QAOA_Experimental}
Another example of heuristic algorithms is quantum annealing, which solves a quadratic binary optimization by mapping it onto the equivalent problem of calculating the ground state configuration of an Ising Hamiltonian.
The latter is then calculated on a specialized simulator.
The idea is that quantum tunneling will explore the complicated energy landscape of this model more efficiently and, therefore, avoid converging into local minima.\cite{Katzgraber2020_Annealing,Hauke2021_ReactionPathways-Annealing}

All these classes of optimization algorithms can be, in principle, applied to biochemical problems such as those mentioned above.
The first attempts to solve the protein folding optimization problem relied on quantum annealing\cite{PerdomoOrtiz2012,Babbush2014} while, more recently, hybrid classical-variational algorithms based on QAOA,\cite{Fingerhuth2018} VQE,\cite{Robert2021} and on quantum random walks\cite{Casares2022_QFold-ProteinFolding} have been proposed.
The efficiency of all these methods has only been explored so far for simplified lattice models of small polypeptides with a few dozen of amino acids.
On the one hand, this may be taken as an indication that these quantum algorithms may be promising candidates for molecular biology applications.
At the same time, it is impossible to assess, based only on this observation, if and which one of them may yield a practical quantum advantage for systems that are out of the reach of classical computers.
Furthermore, going beyond simplified lattice models and applying them to a real force field is a crucial step in order to truly compete with methods based on classical dynamics.
We note that finding the lowest-energy conformation of a polypeptide is not the unique relevant optimization problem in protein science.
For instance, it was proposed in Ref.~\citenum{Hauke2021_ReactionPathways-Annealing} to reformulate the problem finding the most favourable reaction pathways as an optimization problem, and to solve it with quantum-annealing algorithms.

A key conceptual difference between simulation- and optimization-based solutions to the protein folding problem will remain for both classical and quantum computing.
Solving folding as an optimization problem yields only the final structure, corresponding to the lowest (free) energy, but this procedure does not provide any insights into the physical dynamics of the folding process itself.
As illustrated by the Levinthal paradox,\cite{Zwanzig1992_Levinthal} a protein cannot fold simply by randomly sampling all its possible conformations because, in this way, the folding time would be much higher than what is observed experimentally.
The local interactions between amino acids drive, while the protein folds,  the protein toward the lowest-energy conformation, as highlighted by the famous ``folding funnel'' metaphor.
While it is obvious that a quantum mechanical description of these interactions to yield accurate electronic energies that then enter sampling procedures of structural configurations would be desirable, it is highly uncertain whether a quantum machine can accomplish this for such large atomistic structures (let alone the fact that also the role of chaperone proteins and other mechanisms that can assist or steer a folding process may need to be considered).
However, embedding approaches that dissect a quantum system into smaller quantum subsystems, which are then amenable to actual computations, can deliver key concepts to meet this challenge.\cite{Tavernelli2021_Embedding,Vorwerk2022_QuantumComputingEmbeding-Materials,Rossmannek2023_Embedding,Liu2023_BootstrapQC}

Finally, we note that a third strategy to solve the protein folding problem is offered by data-driven approaches.
The three-dimensional structure of many proteins is often available from nuclear magnetic resonance or X-ray crystallography.
It is, therefore, possible to use this enormous amount of experimental data as a training set for a machine-learning model.
A prime example of such an application is the deep neural network AlphaFold.\cite{Jumper2021_AlphaFold}
AlphaFold is a specific machine learning algorithm that can predict the three-dimensional structure of a protein using, as input, its amino acid sequence.
Data-driven approaches can overcome the steep scaling of explicit classical and quantum simulations.
However, since the former make predictions based on previously learned data, the accuracy of machine learning for genuinely new instances depends on its transferability to these unseen input data -- a property that is usually referred to as generalization power, which is extremely hard to estimate \textit{a priori}.

It is therefore not surprising that it was recently shown\cite{Kiersten2021_AlphaFold-Disordered} that data-driven approaches can hardly predict the structure of so-called intrinsically disordered proteins,\cite{Oldfield2014_IntrinsicallyDisorderedProtein} in which only a small fraction of the amino acid sequence is organized into a well-defined three-dimensional structure, while the rest of the protein is inherently dynamical and disordered.
Inferring the structure of these proteins based on data-driven approaches is extremely challenging because their conformation is inherently different from any other protein and inference from data lacking this information breaks down.
Physical modeling such as first-principles simulations does not suffer from these limitations by construction, but
many practical challenges of simulating intrinsically disordered proteins with classical algorithms remain.\cite{Vendruscolo2006_IntrinsicallyDisordered}
We will discuss in the next section whether quantum computing may have the potential to also advance machine learning applications.

\section{Quantum computing for data-driven approaches to molecular biology}
\label{sec:QuantumMachineLearning}

One may argue whether quantum computing has the potential to boost machine-learning algorithms in molecular biology.\cite{Ching2018_DataBiologyReview,Greener2022_DataBiologyReview}
Answering this question is the overarching goal of \emph{quantum machine learning}.\cite{Biamonte2017_QuantumMachineLearning, Verdon2022_QuantumMachineLearning}
In essence, machine learning is a mathematical framework to \emph{learn} a target function
by knowing its value at a given set of points (usually known as \textit{training set}).
The target function
is approximated with a parameterized function, and the parameters are optimized to reproduce its value on the training set.

Quantum computing may boost machine learning in two respects.
First, it can speed up the training process.
Second, it can make the algorithm more expressive in order to be able to predict more complex functions.
The first advantage has been leveraged in deriving the quantum realization of the principal components analysis algorithm\cite{Lloyd2014_QuantumPCA} and of a support vector machine.\cite{Lloyd2014_Quantum-SVM}
Formal guarantees of their quantum speedup over their classical counterpart might be available for these algorithms, but
they require, as the initial step, preparing the qubits in a linear superposition with coefficients that depend on the input data set.
In principle, such a step can be realized with the quantum realization of the random access memory (RAM) architecture.\cite{Giovannetti2008_qRAM}
However, depending on the nature of the input data, this loading phase may become the bottleneck of the overall simulation, therefore eliminating any practical quantum advantage.\cite{Aaronson2015_ReadFinePrint}

Methods that leverage the quantum computer only for boosting the representation power of the machine-learning model do not suffer, in principle, from this limitation because they can carry out the training process classically.
For instance, a neural network approximates the target functions in terms of elementary, non-linear functions -- the so-called \emph{neurons}.
Increasing the number of neurons increases the representation power of the network but, at the same time, makes the training phase more complex.
A quantum circuit (see Figure~\ref{fig:DigitalAnalog}) can be interpreted as a quantum neural network, where the gates are the quantum counterpart of the neurons.
For a given number of neurons, quantum neural networks can be more expressive -- even exponentially, for specific network architectures -- than their classical counterpart.\cite{Cong2019_QuantumConvolutionalNeuralNetworks,Killoran2019_ContinuousVariableQuantumNN,Abbas2021_QuantumNeuralNetworks}
Note that this expressivity boost is not limited to neural networks and has been observed, for instance, for support vector machines.\cite{Lloyd2014_Quantum-SVM,Havlicek2019_KernelQuantumEnhanced,Liu2021_QuantumKernelSpeedup}
However, the training phase, when carried out classically, may itself become hard due to the emergence of so-called Barren plateaus, especially for large data sets.\cite{McClean2018_QNN-Barren,Wiebe2021_QNN-Barren}
This makes the training process itself a computationally hard problem and therefore eradicates a quantum speedup.

Quantum algorithms for solving data-driven problems are not limited to machine learning.
These algorithms can also be used to identify correlations \textit{within} data that are collected experimentally.
For instance, genome sequencing requires reconstructing the nucleobase sequence of an organism based on small portions of the full DNA sequence.\cite{Heather2016_DeNovoAssembly_Review}
By interpreting the DNA segments as vertices of a graph, which are connected if the terminal segments
of their nucleobase sequence overlap, genome sequencing becomes equivalent to the problem of finding a Hamiltonian cycle of the graph.
This problem is known to be NP-hard classically,\cite{Myers2005_FragmentAssembly,Compeau2011_GenomeAssembly} although efficient heuristic classical solutions to it have been developed over the years.\cite{Nagarajan2013_SequenceAssembly,MacLean2009_NextGenerationSequencing}
The heuristic quantum algorithms introduced in Section~\ref{sec:QuantumForClassical} for solving optimization problems can be straightforwardly applied also to this context.

A key take-home message of the discussion so far is that quantum machine learning may become applicable, in the future, only for data-driven quantum algorithms in which a quantum speedup can be observed already for small data sets.
Otherwise, either loading the classical data on quantum hardware or carrying out the classical optimization will become the main bottleneck, nullifying any quantum advantage.
For instance, quantum machine learning algorithms for identifying common patterns between two DNA sequences -- a task known as DNA alignment -- have been proposed in recent years.\cite{Prousalis2019_QuantumPatternRecognition,Sarkar2019_DNAAlignment}
As already mentioned above, the Grover search yields only a quadratic speedup over its classical realization.
With such a limited speedup, a practical advantage will only be observed for problems with very large data sets.\cite{Babbush2021_QuadraticSpeedup}
However, these will be difficult to deal with on currently designed hardware with rather few physical qubits (to be compared to the vast number of bits in a classical computer).
Moreover, it remains to be seen how to experimentally realize a so-called quantum random access memory,\cite{Giovannetti2008_qRAM} which would represent a crucial component for loading a large data set on a quantum computer.
For these reasons, one may not expect quantum algorithms based on the Grover search to represent -- at least in the foreseeable future -- suitable candidates for demonstrating a practical quantum advantage.

Other applications of quantum machine learning algorithms may, in principle, yield a practical quantum advantage in the future.
This is the case of the recently proposed quantum algorithm for DNA sequencing\cite{Sarkar2021_DNADeNovoSequencing,Boev2021_GenomeAssembly} and for identifying the most probable binding site of a transcription factor to DNA.\cite{Li2018}
In both cases, the quantum hardware is leveraged to solve a classically hard optimization problem with heuristic algorithms.
However, being heuristic, the efficiency of these algorithms must be assessed at a case-by-case level.
At this stage of hardware development, with quantum computers composed of at most hundreds of noisy qubits (see Section~\ref{sec:quantum}), these algorithms can be validated only on very small data sets.
It is, therefore, currently extremely hard to predict whether they will yield an advantage when applied to problems that are out of the reach of classical hardware.
This limitation does not only hold for machine learning algorithms, but for any problem where classical data are mapped onto the qubit space, including, among others, the task of finding the lowest-energy configuration of a biological macromolecule given a force field describing the interaction between its atoms.

\section{Conclusions}

The living world is built from atoms and molecules and, therefore, it is clearly governed by quantum physics.
Due to the large size of most biomolecules and the relatively high temperatures of living organisms, classicality emerges via the process of decoherence.
This raises the question in which situations accurate quantum mechanical modeling is required and where effective classical description suffices.
In an in-silico descriptive and predictive context, this question directly affects the computational resources required to simulate biological processes, because full quantum mechanical simulations are affected by the curse of dimensionality when simulated with classical computers.

This article has laid out the prospects of quantum computing coming to rescue, moving from smaller to larger biomolecular structures.
Quantum mechanical simulations are essential when simulating biochemical processes that are accompanied by the creation of new chemical bonds.
Simulating these phenomena has, therefore, the highest prospect of impact  by quantum computers.
However, we also discussed how quantum effects may be crucial also in larger functional biomolecules, where quantum effects shape classical effective descriptions and require a more targeted use of quantum computing at this quantum-classical barrier.

Moving to even larger structures with an effectively cured curse of dimensionality, biocomputations are now faced with solving a still hard, but now entirely classical, computational problem, often in the form of an optimization problem, \textit{e.g.} in protein folding.
Also here, quantum computing might come to supplement classical computers.
From a proven quadratic Grover speedup for exhaustive search strategies to tailored algorithms to be run on fault-tolerant computers, to the heuristic approaches for near-term quantum devices, there is a long list of approaches for this class of problems, which appears not only in biomolecular sciences but in all sciences and industries, ranging from logistics to finance.

Nevertheless, the specifics of the optimization problems in biomolecular sciences require a targeted and in-depth understanding of the type of optimization problem as well as a tailored solution.
The same is true more generally for important classical computational problems in the life sciences, ranging from genomics to patient data from, e.g., imaging or sensory data such as from wearable devices.
Here, it seems possible that quantum computers might become a supplement and accelerator of machine learning methods.
More generally, quantum technologies, such as quantum sensing,\cite{Degen2017_QuantumSensing-Review} might enhance the quality of imaging data of biomolecular structures (see e.g. Ref.~\citenum{polzik-nerve}) and, therefore, improve the quality of the data on which further classical or quantum computations would be based.
In a future scenario, not only the quantum computers but also the quantum sensors might be connected with help of entanglement, thereby harnessing a further increase in resolution (cf. Ref.~\citenum{telescope}).

We emphasize that the first-principles nature of the quantum mechanical basis of the physical modelling approach makes the methods discussed in this work also applicable to other branches of molecular and materials science.
Accordingly, approaches for modelling biochemical processes will also allow for modelling of general chemical reactions.
However, this is true also from a more general perspective of physical modelling or data-driven science.
For instance, approaches for protein simulation will also be applicable in the context of polymer physics.

\section*{Acknowledgments}

The authors are grateful for the generous funding through the 'Quantum for Life Center' funded by the Novo Nordisk Foundation (grant NNF20OC0059939).

\providecommand{\latin}[1]{#1}
\makeatletter
\providecommand{\doi}
  {\begingroup\let\do\@makeother\dospecials
  \catcode`\{=1 \catcode`\}=2 \doi@aux}
\providecommand{\doi@aux}[1]{\endgroup\texttt{#1}}
\makeatother
\providecommand*\mcitethebibliography{\thebibliography}
\csname @ifundefined\endcsname{endmcitethebibliography}
  {\let\endmcitethebibliography\endthebibliography}{}

\end{document}